\documentclass[10pt,conference]{IEEEtran}
\usepackage{cite}
\usepackage{amsmath,amssymb,amsfonts}
\usepackage{graphicx}
\usepackage{textcomp}
\usepackage{xcolor}
\usepackage{tikz}
\usepackage{pifont}
\usepackage[hyphens]{url}
\usepackage{fancyhdr}
\usepackage{hyperref}
\usepackage{setspace}
\usepackage{algorithm}
\usepackage{hyperref}
\usepackage{algorithmicx}
\usepackage{algpseudocode}
\usepackage{mdframed}
\usepackage{media9}
\usepackage[most]{tcolorbox} 
\usepackage{booktabs}
\usepackage{mathbbol}
\usepackage{media9}
\usepackage{makecell}
\usepackage{multirow}
\usepackage{marvosym}
\hypersetup{
  colorlinks=true,
  linkcolor=blue,
  urlcolor=blue,
  citecolor=blue
}

\tcbset{observationbox/.style={
  colframe=black, colback=gray!5,
  boxrule=0.5pt, arc=1pt,
  left=4pt, right=4pt, top=0.2pt, bottom=0.2pt,
  fonttitle=\bfseries
}}

\newcommand{\hpcayear}{2026}

\newcommand{\hpcasubmissionnumber}{132}
\title{TEMP: A Memory Efficient Physical-aware Tensor Partition-Mapping Framework on Wafer-scale Chips}

\def\hpcacameraready{} 

\newcommand\hpcaauthors{Huizheng Wang$^{\dagger*}$, Taiquan Wei$^{\dagger*}$, Zichuan Wang$^{\dagger}$, Dingcheng Jiang$^\dagger$, Qize Yang$^\dagger$, Jiaxin Liu$^\dagger$, Jingxiang Hou$^\dagger$, \\ 
Chao Li$^\ddagger$, Jinyi Deng$^\dagger$\textsuperscript{\Letter}, Yang Hu$^\dagger$\textsuperscript{\Letter}, Shouyi Yin$^\dagger$$^\S$}
\newcommand\hpcaaffiliation{$^\dagger$School of Integrated Circuits, BNRist, Tsinghua University, Beijing, China, 100084 \\
$^\ddagger$School of Computer Science and Engineering, Shanghai Jiao Tong University, Shanghai, China, 200240 \\
$^\S$Shanghai Artificial Intelligence Laboratory, Shanghai, China, 200433}
\newcommand\hpcaemail{\textsuperscript{\Letter}Corresponding author, dengjinyi@mail.tsinghua.edu.cn; hu\_yang@tsinghua.edu.cn}

\author{
  \ifdefined\hpcacameraready
    \IEEEauthorblockN{\hpcaauthors{}}
      \IEEEauthorblockA{
        \hpcaaffiliation{} \\
        \hpcaemail{}
      }
  \else
    \IEEEauthorblockN{\normalsize{HPCA \hpcayear{} Submission
      \textbf{\#\hpcasubmissionnumber{}}} \\
      \IEEEauthorblockA{
        Confidential Draft \\
        Do NOT Distribute!!
      }
    }
  \fi 
}

\fancypagestyle{camerareadyfirstpage}{%
  \fancyhead{}
  
  \fancyhead[C]{
    \ifdefined\aeopen
    \parbox[][12mm][t]{13.5cm}{\hpcayear{} IEEE International Symposium on High-Performance Computer Architecture (HPCA)}    
    \else
      \ifdefined\aereviewed
      \parbox[][12mm][t]{13.5cm}{\hpcayear{} IEEE International Symposium on High-Performance Computer Architecture (HPCA)}
      \else
      \ifdefined\aereproduced
      \parbox[][12mm][t]{13.5cm}{\hpcayear{} IEEE International Symposium on High-Performance Computer Architecture (HPCA)}
      \else
      \parbox[][0mm][t]{13.5cm}{\hpcayear{} IEEE International Symposium on High-Performance Computer Architecture (HPCA)}
    \fi 
    \fi 
    \fi 
    \ifdefined\aeopen 
      \includegraphics[width=12mm,height=12mm]{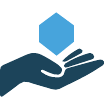}
    \fi 
    \ifdefined\aereviewed
      \includegraphics[width=12mm,height=12mm]{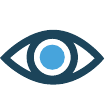}
    \fi 
    \ifdefined\aereproduced
      \includegraphics[width=12mm,height=12mm]{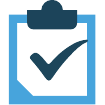}
    \fi
  }
  \fancyfoot[C]{}
}
\begin{document}
\maketitle

\ifdefined\hpcacameraready 
  \thispagestyle{camerareadyfirstpage}
  \pagestyle{empty}
\else
  \thispagestyle{plain}
  \pagestyle{plain}
\fi

\newcommand{\hpcaheight}{0mm}
\ifdefined\eaopen
\renewcommand{\hpcaheight}{12mm}
\fi


\begin{abstract}
Large language models (LLMs) demand significant memory and computation resources. Wafer-scale chips (WSCs) provide high computation power and die-to-die (D2D) bandwidth but face a unique trade-off between on-chip memory and compute resources due to limited wafer area. Therefore, tensor parallelism strategies for wafer should leverage communication advantages while maintaining memory efficiency to maximize WSC performance. However, existing approaches fail to address these challenges.

To address these challenges, we propose the tensor stream partition paradigm (TSPP), which reveals an opportunity to leverage WSCs' abundant communication bandwidth to alleviate stringent on-chip memory constraints. However, the 2D mesh topology of WSCs lacks long-distance and flexible interconnects, leading to three challenges: 1) severe tail latency, 2) prohibitive D2D traffic contention, and 3) intractable search time for optimal design.

We present TEMP, a framework for LLM training on WSCs that leverages topology-aware tensor-stream partition, traffic-conscious mapping, and dual-level wafer solving to overcome hardware constraints and parallelism challenges. These integrated approaches optimize memory efficiency and throughput, unlocking TSPP's full potential on WSCs. Evaluations show TEMP achieves 1.7$\times$ average throughput improvement over state-of-the-art LLM training systems across various models.
\end{abstract}

\section{Introduction}\label{sec:introduction}
\footnotetext[1]{These authors contributed equally to this work.
}
Large language models (LLMs) have emerged as pivotal components in advancing artificial intelligence (AI) \cite{brown2020language,radford2018improving,liu2021post,lan2019albert,liu2019roberta,yuan2021tokens,song2018situ}. The LLM scaling law \cite{zhang2024scaling} highlights model size as a key performance driver, exemplified by over $450\times$ model size increase from GPT-2 \cite{radford2019language} to DeepSeek \cite{deepseekai2024deepseekv3technicalreport}. 
Unfortunately, this rapid growth imposes significant demands on computation resources, making current monolithic devices increasingly difficult to provide sufficient transistor density for LLM training \cite{han2021pre}.

Wafer‐scale chip (WSC) design, enabled by advanced packaging technologies like TSMC’s CoWoS \cite{huang2021wafer}, offers a promising solution to mitigate these issues, by integrating numerous dies on a wafer-scale substrate. Compared to current board-level DGX systems,  WSCs typically can offer $6\times$ higher D2D bandwidth and $5\times$ lower latency \cite{talpes2022dojo,la2020cerebras,hu2024wafer}, benefiting from the finer and higher density interconnect pitches.

However, different from current ASIC designs \cite{moon2025t,wang2025beta,kim2025slim,wang2025bitstopper,dong202528nm,wang2023efficient,wang2025mcbp,wang2025lapa}, chiplet \cite{li2022spacx,sudarshan2024eco,sharma2025heterogeneous,shan2022architecture} and DGX-base designs \cite{xie2021fast,gawande2020scaling,udomchoksakul2022gpu}, WSC-based systems face a unique architecture trade-off that both memory and compute resources are limited by the wafer’s area, typically 40,000mm$^2$. In other words, increasing on-wafer memory capacity comes at the cost of compute resources. \textbf{Therefore, LLM training on WSC-based systems necessitates the adoption of a memory-efficient tensor parallelism strategy.}

\begin{figure}[t]
\centering
\includegraphics[width=\linewidth]{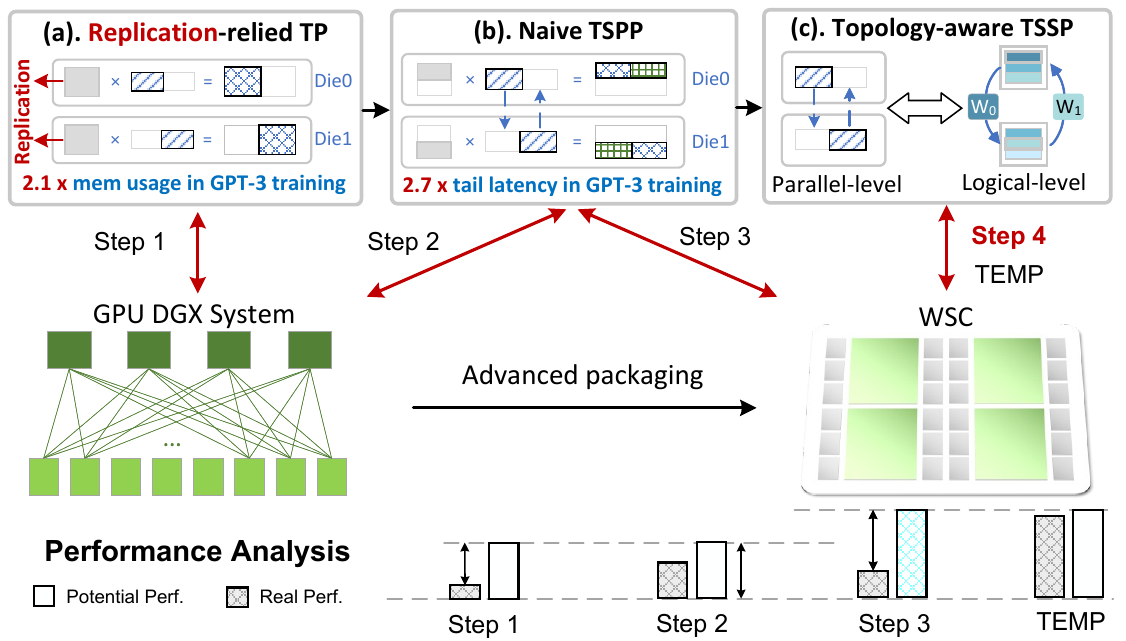}\vspace{-6mm}
\caption{Illustrations for co-design features of the TEMP framework.}
\label{fig:intro}
\end{figure}
 
However, by re-examining existing tensor partition frameworks \cite{shoeybi2019megatron,beaumont2021efficient,zheng2022alpa,zhao2023pytorch}, we identify that their parallelisms involve redundant tensor replication. As depicted in Fig.\ref{fig:intro} (a), although the
weights are partitioned and stored separately across two dies, the activations remain replicated on both. 

To tackle the memory inefficiency issue, inspired by the distributed GEMM algorithms \cite{cannon1969cellular,van1997summa,schatz2016parallel}, we design a stream-style tensor partition mechanism, named TSPP. As shown in Fig. \ref{fig:intro}(b), TSPP partitions both input and weight tensors into non-overlapping sub-tensors, performs fine-grained sub-computations. While computing on a subset of these sub-tensors, the remaining sub-tensors are swiftly exchanged. TSPP offers two advantages: it eliminates tensor replication and enables the overlap of communication with computation.

\textbf{However, we identify that naively applying TSPP on WSCs is sub-optimal due to severe tail latency}, as shown in the step 3 of Fig.\ref{fig:intro}. This issue occurs because, when mapped onto physical dies, TSPP requires a ring configuration. If the dies are arranged linearly or in a non-ring configuration, the transmission latency between the first and last dies will far exceed that between adjacent dies, resulting in tail latency and reduced compute utilization.

Naturally, one may naturally consider physically adding a torus link. However, this is impractical to achieve at WSCs, where the side length typically exceeds 190 mm. Once the die-to-die transmission distance surpasses 50\,mm, the bit error rate increases by up to 10$^8\times$ \cite{chen2023floorplet,pal2020pathfinding,yang2025pd}. As a result, forward error correction (FEC) becomes necessary, increasing the transmission latency to 210 ns \cite{sella2018fec}, which is $14\times$ higher than that in a normal scenario. This substantial time overhead in turn further exacerbates the tail latency.

To this end, we design a topology-aware TSPP, termed TATP, which features coordinated optimization across both parallel and logical communication levels. By incorporating logical-level communication optimizations, TATP can achieve higher performance on WSCs, without exacerbating tail latency, as illustrated in the step 4 of Fig. \ref{fig:intro}.

Unfortunately, it is still far from trivial to implement TATP on WSCs. As depicted in Fig. \ref{fig:framework} (a), key challenges lies in: \textbf{(1) Prohibitive traffic contention}. This is because when TATP collaborates with existing parallel schemes, the lack of a unified global mapping leads to communication path conflicts between different parallelisms, causing congestion. \textbf{(2) Intractable search time for the optimal parallel configuration.} This is because new TATP extends the parallelization space, while the numerous integrated dies on the WSC dramatically expand the design space for feasible mappings.

To address the above challenges, we design TEMP, a framework that systematically integrates TATP with existing parallel schemes and accelerates the search for the optimal parallel configuration on WSCs. In summary, TEMP incorporates the following key innovations:

\begin{figure}[t]
\centering
\includegraphics[width=\linewidth]{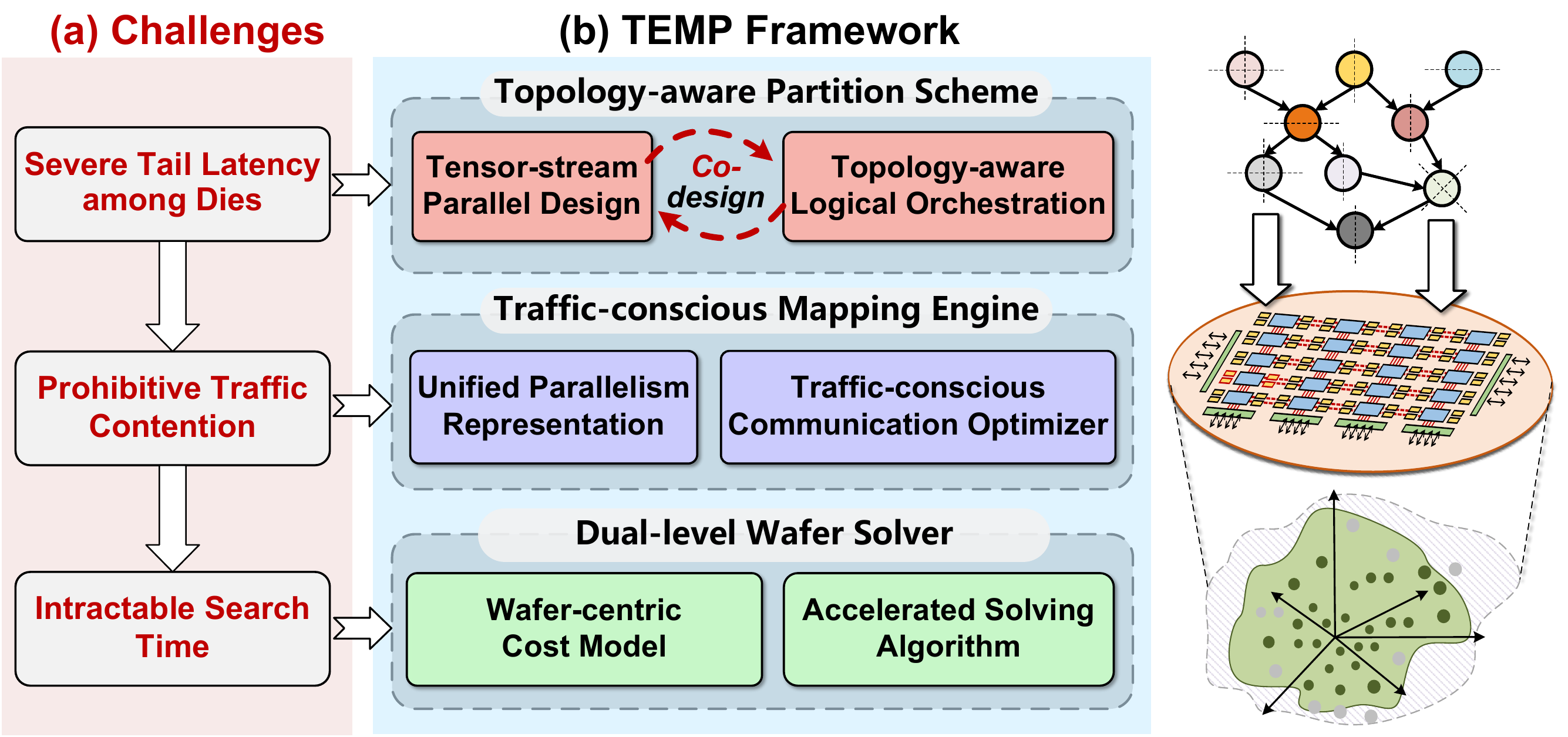}\vspace{-6mm}
\caption{Challenges and the TEMP solution for tensor-stream execution on wafer-scale chips. 
(a) Implementation challenges. 
(b) Key contributions and components of the TEMP framework.}
\label{fig:framework}
\end{figure}

\textbf{1)} We systematically analyze the architectural characteristics of wafer-scale systems, including their physical scale and communication constraints. Our analysis shows that a naive TSPP design is infeasible at the wafer scale due to the lack of dedicated physical torus links, which results in excessively long communication paths and severe reliability issues.


\textbf{2)} We propose the Topology-Aware Tensor Partitioning (TATP) scheme, which jointly co-designs tensor partitioning and logical execution orchestration with explicit awareness of the wafer-scale topology. By aligning communication patterns with the underlying physical interconnect and eliminating long-hop transmissions, TATP effectively exploits high-bandwidth die-to-die links to improve computation resource utilization while maintaining strict memory efficiency.


\textbf{3)} We introduce the Traffic-Conscious Mapping Engine (TCME), which provides a unified representation of various parallel execution schemes through a dedicated mathematical encoding. TCME then leverages a traffic-aware communication optimizer to intelligently map parallel executions, dynamically reducing communication congestion by optimizing data flow paths and balancing network load.
    

\textbf{4)} We propose a Dual-Level Wafer Solver (DLWS), which combines a wafer-centric cost model with a customized dynamic programming algorithm. This integration enables efficient exploration of optimal parallel configurations across the vast design space of WSCs, significantly reducing the search complexity while ensuring high-quality solutions.


\textbf{5)} Comprehensive evaluation across multiple LLM models shows that TEMP can achieve an average overall speedup of $1.7\times$ and $1.9\times$ higher power efficiency compared to state-of-the-art (SOTA) LLM training frameworks, demonstrating its robust performance across various configurations.


\begin{figure}[t]
\centering
\includegraphics[width=\linewidth]{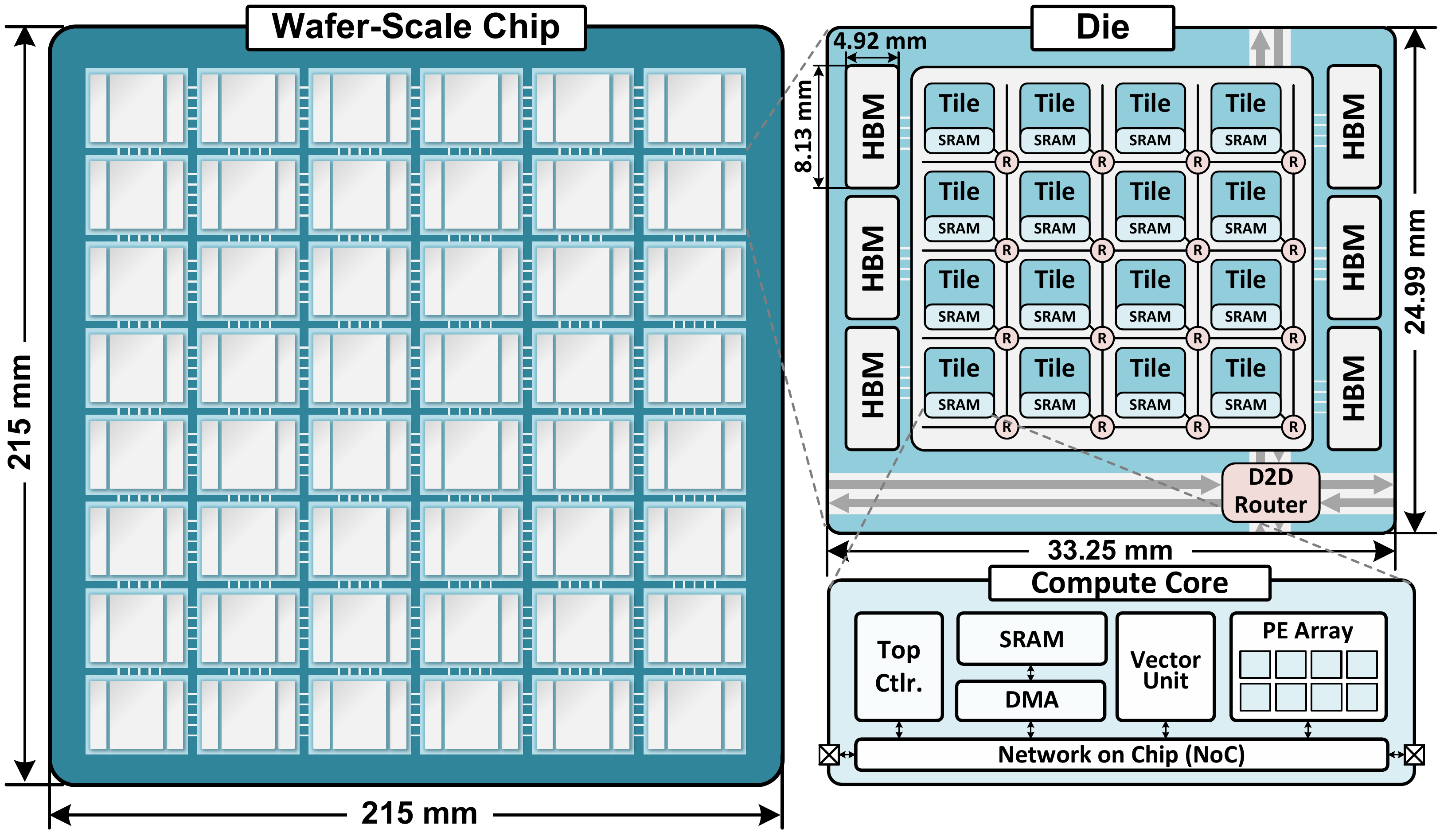}\vspace{-6mm}
\caption{Hardware configurations of WSCs.}
\label{fig:Hardware configurations of WSCs}
\end{figure}

\section{Background}\label{sec:background}

\subsection{Distributed DNN Training}

To address the substantial computational and memory demands of training and inference with large-scale models, a variety of parallelization strategies have been developed. These strategies exploit different forms of parallelism to improve scalability and efficiency, including data parallelism (DP), tensor parallelism (TP), sequence parallelism (SP), context parallelism (CP), and pipeline parallelism (PP). 

In DP~\cite{abadi2016tensorflow,goyal2017accurate,xing2015petuum,li2020pytorch}, each worker maintains a full replica of the model, processes a distinct dataset subset (a.k.a, mini-batch), and periodically synchronizes gradients to ensure model consistency. TP~\cite{dean2012large,wang2019supporting,wang2024primepar,shoeybi2019megatron} partitions tensors along specific dimensions, and distributes the resulting tensor slices across devices. SP~\cite{korthikanti2023reducing} alleviates memory bottlenecks caused by long input sequences by partitioning activations along the sequence dimension. While SP requires communication (e.g., all-gather) to reconstruct the full sequence for attention calculations, it effectively distributes the activation memory footprint. CP~\cite{nvidia_CP,yang2024context,li2021sequence} addresses the challenge of long-context inference by partitioning the attention context window across devices. To further reduce communication overhead and balance workload, several dedicated attention optimizations have been proposed~\cite{liu2023ring,brandon2023striped,fang2024usp}. Each device stores and processes a portion of the KV Cache, computes attention locally, and aggregates partial results, enabling inference with virtually unbounded context lengths by scaling the number of devices. PP~\cite{huang2019gpipe,narayanan2021memory,narayanan2019pipedream} partitions the model into multiple stages, each assigned to a different device, with inter-stage communication efficiently managed via point-to-point links, minimizing overhead. 

Despite the diversity of existing parallelization strategies, not all of them are equally suitable for wafer-scale systems. This work focuses on optimizing TP, DP, in conjunction with SP and CP, while excluding PP, driven by the following considerations: First, although PP can reduce communication traffic, it often incurs substantial pipeline bubbles, leading to degraded training throughput. Second, on WSCs with high D2D bandwidth, employing PP is suboptimal, as it fails to fully exploit the abundant communication capability provided by the underlying architecture.

\begin{figure}[t]
\centering
\includegraphics[width=\linewidth]{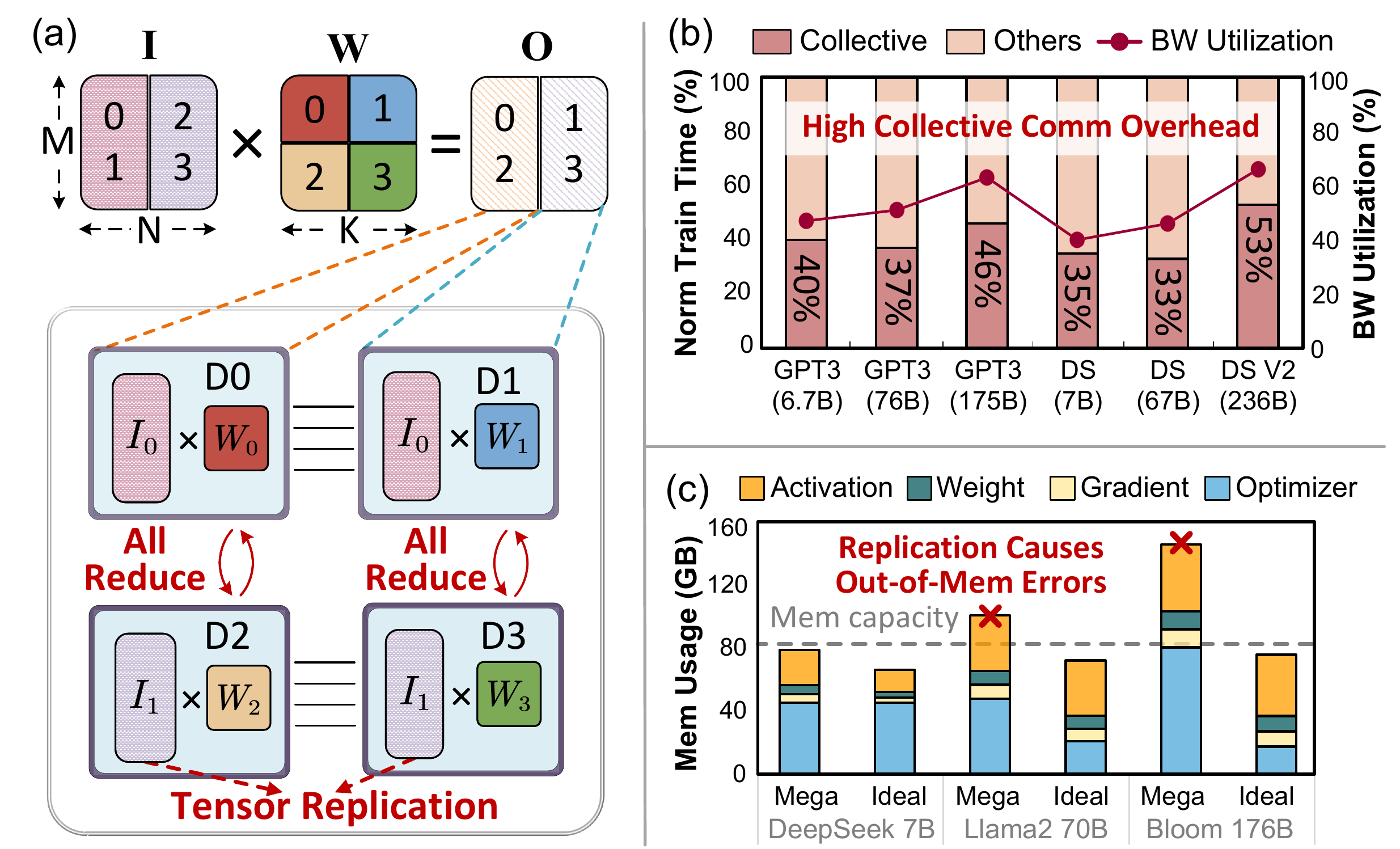}\vspace{-5mm} 
\caption{(a) Illustration of 4-way tensor parallelism, where colors denote tensor partitions and numbers indicate device IDs. (b) Training time breakdown for GPT-3 and DeepSeek models with Megatron-LM. (c) Memory overhead of Megatron-LM compared to an ideal baseline. The dashed red line marks the WSC's per-die memory capacity.}
\label{fig:Traditional Parallelism}
\end{figure}

\subsection{Wafer-scale Chip (WSC)}\label{sec:waferscale chip}

WSCs demonstrate strong potential for high-density compute and interconnect resources by leveraging advanced packaging techniques \cite{bajwa2018demonstration,hou2017wafer,iyer2019silicon,berru2023demonstration}, such as TSV \cite{jiao2024low, guo2022review, lau2023recent} and chip-on-wafer (CoW) \cite{huang2021wafer, hu2023cowos, yang2024signal} processes. Existing WSC implementations generally follow two design approaches. Cerebras \cite{Cerebras2022Cerebras,lie2024inside,Cerebrastraining} adopts a monolithic wafer design, relying on offset exposures and proprietary interconnects to integrate dies on a monolithic wafer. In contrast, other efforts \cite{talpes2022dojo,shih2025sow,pal2021designing,pal2019architecting,bajwa2018demonstration} employ chiplet-based heterogeneous integration, where compute and memory dies are fabricated separately and bonded at the wafer scale. The latter approach enhances design flexibility and manufacturing yield by leveraging known-good-die (KGD) techniques. Owing to its generality and yield advantages, this work focuses on heterogeneously integrated WSCs.

\textbf{Wafer-level configuration.} The WSC architecture is organized into three hierarchical levels: wafer, die, and core. As depicted in Fig. \ref{fig:Hardware configurations of WSCs} left, the wafer integrates a 6$\times$8 array of dies arranged in a 2D-mesh, with dedicated IO dies placed along the periphery to support external communication. This design enables 4 TB/s bandwidth across a 46225\,mm$^2$ wafer while maintaining wafer-scale production feasibility.

\textbf{Die-level configuration.} Each die integrates computing cores, high-bandwidth memory (HBM), and a network-on-chip (NoC) within a 24.99 mm × 33.25 mm footprint. The hierarchical NoC consists of intra-die routers that connect computing cores and die-to-die (D2D) routers that interface with memory controllers to support cross-die communication, as illustrated in Fig. \ref{fig:Hardware configurations of WSCs} right. Each HBM stack provides up to 0.8 TB/s of memory bandwidth, enabled by the high-speed HBM3 interface.

\textbf{Core-level configuration.} The compute cores are architecturally optimized for computational workloads, featuring a top controller that dispatches pre-configured instruction streams and orchestrates data transfers between SRAM and DRAM. DMA and NoC collectively manage inter-core communications, ensuring high-throughput data movement. PE arrays and vector units efficiently perform GEMM/GEMV and vector/scalar operations, respectively.

\section{Motivation}\label{sec:motivation}

\subsection{Inefficiency of Current Tensor Partition}\label{subsec:current_tensor_partition}
Tensor partitioning has been extensively explored as a core technique for neural network parallelization to accelerate computation and amortize memory overhead \cite{coates2013deep,shoeybi2019megatron,kim2016neurocube,zheng2022alpa,zhao2023pytorch,dean2012large,yang2016systematic,wang2024primepar}. Despite this progress, we observe that existing partitioning strategies leave significant portions of the design space unexplored, leading to suboptimal hardware utilization. Fig.~\ref{fig:Traditional Parallelism}(a) illustrates this inefficiencyby examining Megatron-LM’s execution strategy for a linear layer \cite{shoeybi2019megatron,narayanan2021efficient}. The weight matrix $W$ is partitioned into 2$\times$2 blocks along the (N,K) dimensions and distributed across four devices. The input activation $I$ is halved along the N dimension and \textbf{replicated} across device pairs. As a result, devices D0 and D2 each operate on $W_0$, $W_2$ respectively, and must perform an \textbf{all-reduce} communication to aggregate their partial sums, incurring extra data traffic. 

Notably, this communication traffic or tensor replication overhead is not an artifact of the implementation but an inherent consequence of enforcing stationary tensor placement. Maintaining tensor stationarity constrains the mapping of computation to hardware, inducing redundant data movement and synchronization. We refer to this fundamental limitation as the \textit{Stationary Tensor Partitioning Mechanism}.


\begin{figure}[t]
\centering
\includegraphics[width=0.97\linewidth]{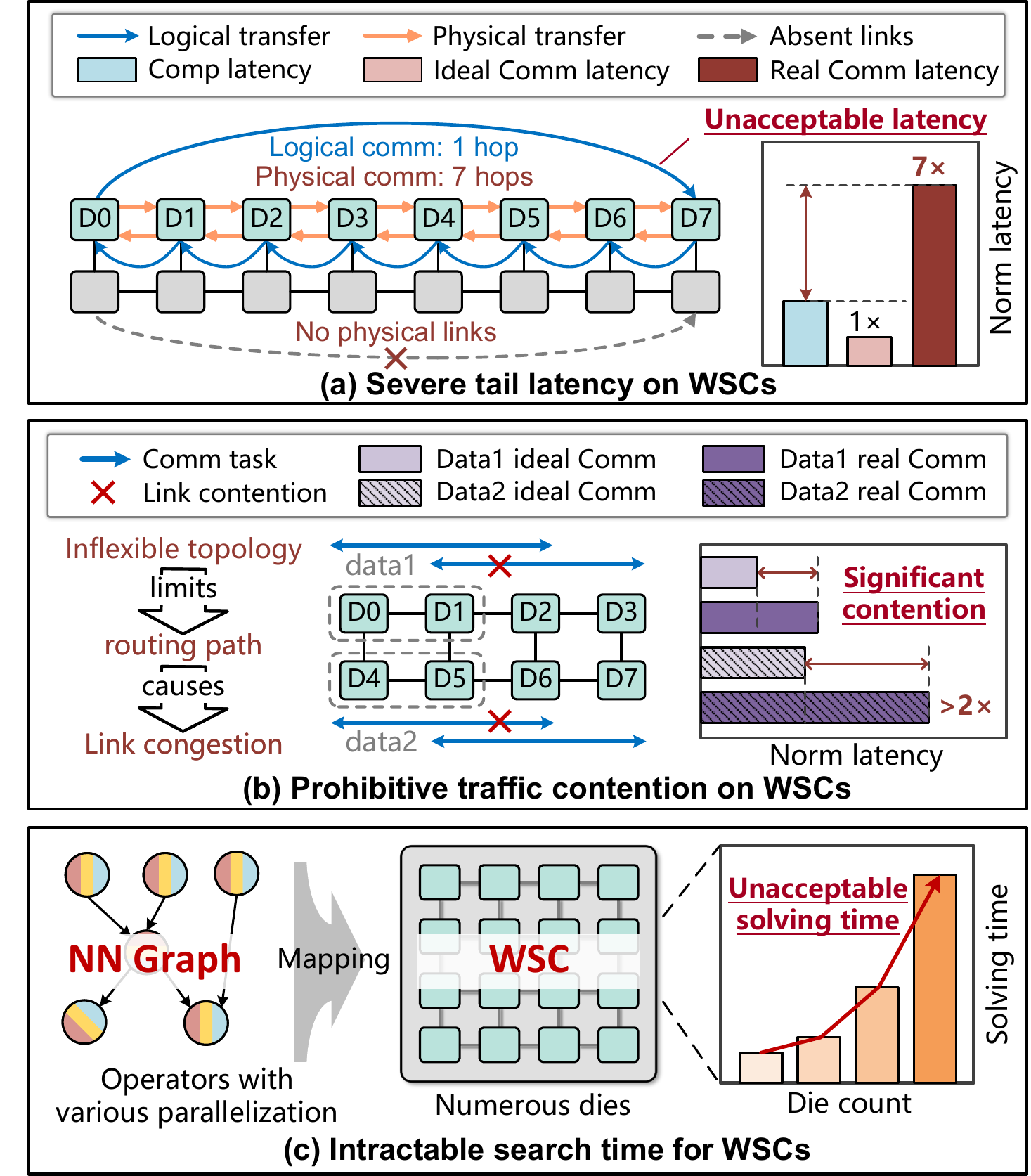}
\caption{Challenges for deploying TSPP on WSC. $D_i$ means Die$i$ on a WSC.}
\label{fig:motivation fig2}
\end{figure}

Our analysis further quantifies the extent to which such tensor replication and collective communication degrade the WSC training efficiency. As shown in Fig.~\ref{fig:Traditional Parallelism}(b), collective communication accounts for 40\% of the total training time, while the D2D bandwidth utilization remains suboptimal. staying below 55\% due to idle D2D interconnects during computation. This inefficiency is exacerbated by the coupling of communication with stationary tensor placement. 

Additionally, tensor replication wastes valuable on-wafer memory, directly limiting the WSC's ability to train larger models. For instance, as shown in Fig.~\ref{fig:Traditional Parallelism}(c), training the Llama3 70B model with TP=8 and DP=4 results in extensive activation replication, leading to a 1.4$\times$ increase in memory consumption and frequent out-of-memory (OOM) errors. Without such replication, the WSC can successfully train the model with no memory overflows. Furthermore, we can notice that as the model size increases, memory demand grows sharply. This highlights the importance of enabling larger-model training under the fixed physical capacity of WSCs.

To address the memory and communication inefficiencies inherent in the current tensor partition strategy, we propose the \emph{Tensor Stream Partition Parallelism (TSPP)}. The core idea of TSPP is to partition tensors into non-overlapping sub-blocks to eliminate redundant tensor replication, utilize high-bandwidth interconnects to swiftly exchange sub-block tensors, and overlap communication latency with fine-grained computation. In this way, TSPP not only eliminates unnecessary memory usage but also alleviates collective communication, by enhancing D2D interconnect utilization. It is noteworthy that the swift exchange of sub-tensors requires substantial interconnect bandwidth from the underlying physical platform. The abundant D2D interconnect bandwidth provided by WSCs makes them a promising platform for deploying TSPP.

\subsection{Challenges of Deploying TSPP on WSCs}\label{sec:physical constraints}
While WSCs provide abundant aggregate interconnect bandwidth, offering a strong theoretical foundation for implementing TSPP, practical deployment faces nontrivial challenges. In particular, constraints imposed by the physical topology, such as limited bisection bandwidth, non-uniform link distribution, and routing contention—can significantly hinder the realization of TSPP’s potential performance gains. 



\vspace{2mm}

\hspace{-3.5mm}
\fbox{%
  \begin{minipage}{0.96\linewidth}
  \vspace{1pt}
\textbf{(Challenge 1)} Severe tail latency due to physical interconnect constraints.
  \vspace{1pt}
  \end{minipage}
}

\vspace{2mm}


As discussed in \S\ref{subsec:current_tensor_partition} and depicted in Fig. \ref{fig:motivation fig2}(a), TSPP relies on a logical ring for data exchange. However, on WSC interposers, signal integrity (SI) sharply degrades beyond 50 mm \cite{yang2025pd}, which precludes direct long-distance or diagonal D2D links \cite{chen2023floorplet,pal2020pathfinding,sella2018fec}. If one ignores these limits and directly implements TSPP on WSCs, the resulting communication will traverse multi-hop paths, leading to severe tail latency \cite{rashidi2024fred}. For instance, as shown in Fig.~\ref{fig:motivation fig2}(a), deploying TSPP across Dies 0–7 creates apparent single-hop transfers (blue arrows), but Die 0 and Die 7 actually require seven physical hops (yellow arrows), while other logical neighbors incur just one. This 7$\times$ communication disparity inflates tail latency, negating TSPP’s benefits. Moreover, although on-wafer D2D links deliver thousands of GB/s, they require large transfer granularities, typically tens to hundreds of megabytes \cite{won2024tacos,laskar2024enhancing,ueno2019exhaustive}, to achieve peak efficiency. Such transfer sizes are comparable to the activation or weight tensor in modern LLM operators \cite{touvron2023llama1,touvron2023llama,zhang2022opt,brown2020language}. This makes subdividing tensors into sufficient pipeline chunks infeasible, leading to link underutilization and exacerbated tail latency. Solving this requires topology-aware tensor partitioning.

\begin{figure}[t]
\centering
\includegraphics[width=0.99\linewidth]{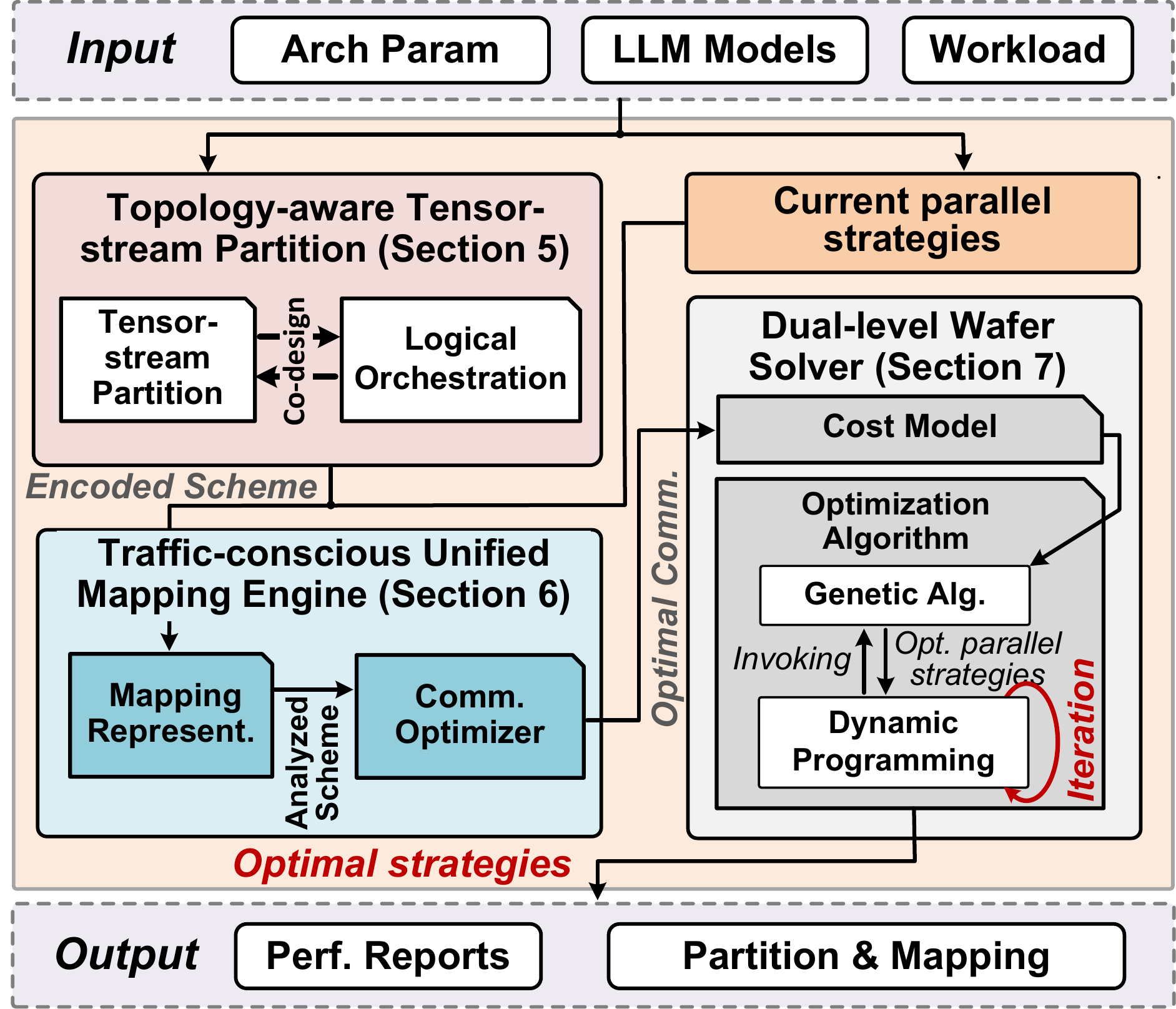}\vspace{-5mm}
\caption{End-to-end overview of the proposed TEMP framework.}
\label{fig:TEMP framework}
\end{figure}


\vspace{2mm}

\hspace{-3.5mm}
\fbox{%
  \begin{minipage}{0.96\linewidth}
  \vspace{1pt}
\textbf{(Challenge 2)} Prohibitive traffic contention due to inflexible topology.
  \vspace{1pt}
  \end{minipage}
}

\vspace{2mm}

The 2D mesh topology fundamentally limits D2D interconnect flexibility \cite{he2025waferllm,avresky2014congestion}, reducing routing path diversity\cite{bani2011performance}. As a result, large-volume communication—often involving hundreds of megabytes during LLM training \cite{rashidi2024fred,li2024understanding}, is prone to severe traffic contention. For example, as illustrated in Fig.~\ref{fig:motivation fig2}(b), data 1 is replicated on Dies 0-1, while data 2 is replicated on Dies 4-5. Assuming that both datasets are transferred to dies on the right half (e.g. blue arrow from Die 0 to Die 2), routing paths from Dies 0 to 2 and 1 to 3 must share the link between Dies 1-2, resulting in traffic contention (red cross mark). This contention increases transfer latency by more than $2\times$ compared to the contention-free case. Consequently, a traffic-conscious mapping engine is essential for WSCs.


\vspace{2mm}

\hspace{-3.5mm}
\fbox{%
  \begin{minipage}{0.96\linewidth}
  \vspace{1pt}
\textbf{(Challenge 3)} Intractable search time due to 
huge design space and the absence of explicit WSC modeling.
  \vspace{1pt}
  \end{minipage}
}

\vspace{2mm}

As illustrated in Fig. \ref{fig:motivation fig2}(c), mapping operators across dies on a WSC under multidimensional parallelism generates an enormous search space. 
Given $N$ dies and $m$ operators, the search complexity grows as $\Omega(N^m)$, rendering the identification of an optimal mapping strategy prohibitively expensive. For instance, when using Integer Linear Programming (ILP) algorithms on an Intel Xeon E5-2686 v4 (Broadwell) CPU, determining the optimal parallelization strategy for GPT-3-76B across 64 dies consumes approximately 40 hours \cite{zheng2022alpa}. Scaling the system to 80 dies further exacerbates this challenge, increasing the search time to over 1000 hours. This prohibitive search time necessitates the development of more efficient and scalable search algorithms.

\section{Overview of the TEMP Framework}
As discussed above, deploying TSPP on WSCs is fundamentally constrained by the wafer’s rigid physical topology, which manifests in severe tail latency, prohibitive inter-die traffic contention, and an intractably large design space for parallelization and mapping.


To systematically address these challenges, we present TEMP, a holistic co-exploration framework that jointly optimizes tensor partitioning and execution mapping for WSC-based LLM training. As depicted in Fig. \ref{fig:TEMP framework}, TEMP takes architectural specifications, LLM model characteristics, and workload specifications as inputs. To address the challenges posed by the rigid network topology, TEMP introduces a \textbf{Topology-aware Tensor-stream Partition Scheme} (\S\ref{sec:TSP scheme}), which aligns tensor-stream execution with the physical interconnect constraints of WSCs and enables efficient deployment of TSPP. Building on this foundation, a \textbf{Traffic-conscious Unified Mapping Engine} (\S\ref{sec:mapping engine}) integrates TSPP with existing parallelization strategies and systematically optimizes communication patterns to mitigate network contention. Finally, a \textbf{Dual-level Wafer Solver} (\S\ref{sec:wafer solver}) employs a wafer-customized cost model together with a dual-level search algorithm to efficiently explore the expanded design space and identify optimal parallelism configurations.

\begin{figure}[t]
\centering
\includegraphics[width=\linewidth]{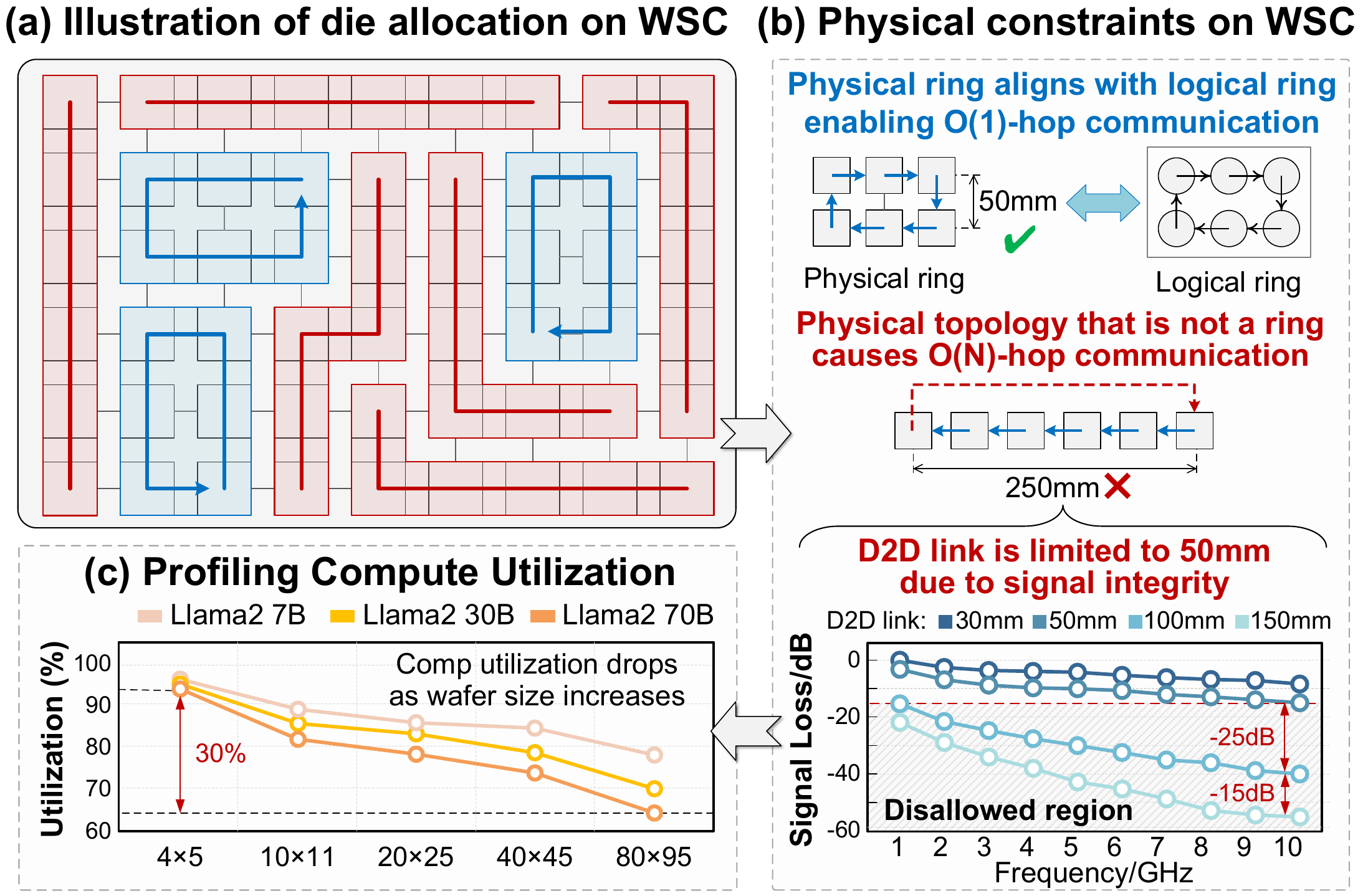}\vspace{-5mm} 
\caption{Motivation for TATP. (a) Parallel task mapping on a wafer-scale mesh: contiguous physical ring groups (blue, efficient) versus non-contiguous ring groups (red, inefficient). (b) The large physical size of the WSC prevents the deployment of long-distance interconnects due to signal integrity limitations. (c) Multi-hop communication induced by non-contiguous ring groups exacerbates tail latency, resulting in a pronounced reduction in overall computation utilization. }
\label{fig:TATP motivation}
\end{figure}

\begin{figure*}[t]
\centering
\includegraphics[width=\linewidth]{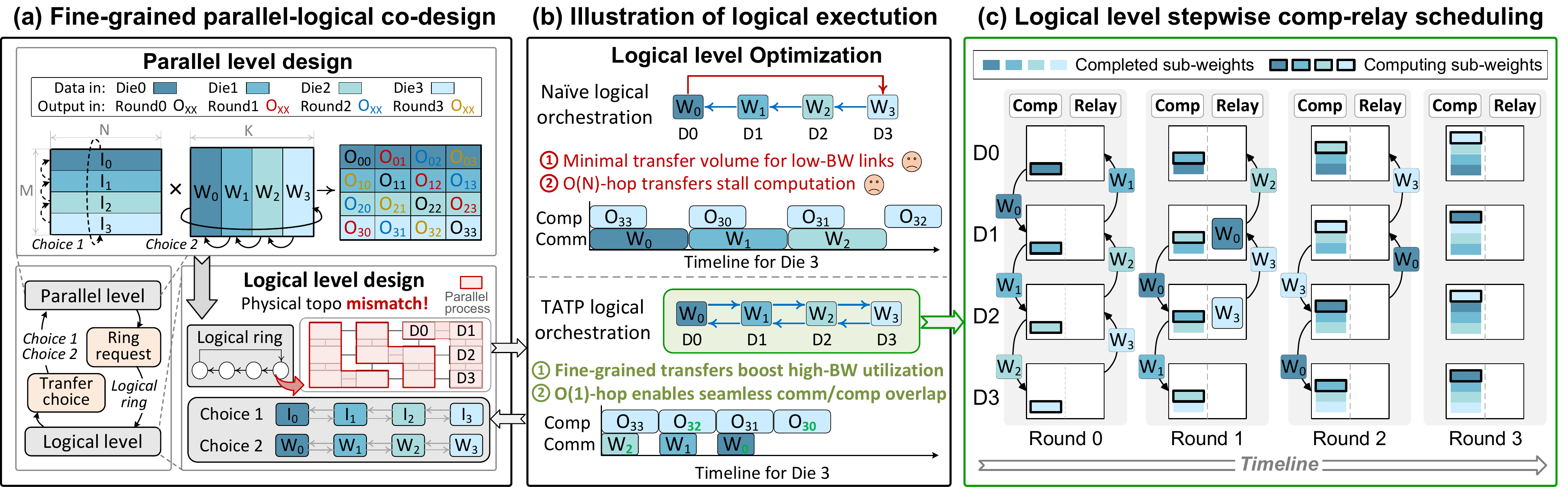}\vspace{-5mm}
\caption{Parallel and logical co-design of TATP. (a) Two-level co-design: Fine-grained parallelization and logical orchestration. (b) Logical level execution optimization in TATP. (c) Stepwise compute-and-relay scheduling of logical execution design.}
\label{fig:Tensor flow partition}
\end{figure*}

\section{Topology-aware Tensor-stream Partition}\label{sec:TSP scheme}
In this section, we introduce Topology-Aware Tensor-Stream Partitioning (TATP), a topology-aware realization of TSPP tailored for WSCs. TATP jointly co-designs tensor partitioning at the parallel level and execution orchestration at the logical level to eliminate severe tail latency under the physical constraints of WSCs.


We begin with a set of motivating examples that highlight why TATP is essential for efficient execution on WSCs. As shown in Fig.~\mbox{\ref{fig:TATP motivation}}(b), realizing a logical ring communication on a wafer requires a contiguous physical ring of adjacent dies. Without such a contiguous physical ring, communication must traverse multiple hops, incurring $O(N)$-hop latency. This constraint is rooted in the physical limitations of 2.5D interposer-based integration, which lacks support for long-range cross-die links \cite{yang2025pd}. As depicted in Fig.\ref{fig:TATP motivation} (b) below, while short ($<$50\,mm) interconnects can tolerate signal loss (e.g., $<$16\,dB), signal integrity deteriorates rapidly once interconnect length exceeds 100–150 mm, making such links unreliable for on-wafer communication. As a result, practical die-to-die interconnects on WSCs are restricted to adjacent dies. In contrast, without the D2D interconnects on 2.5D interposer, traditional GPU clusters leverage switches to create all-to-all topologies, enabling physical rings between arbitrary GPUs via switch routing \cite{lei2025flash}. Therefore, on wafer-scale systems, considering this constraint is critical. 

Further, as our characterizations in Fig.~\ref{fig:TATP motivation}(a), consider a 6$\times$9 die array (54 dies) with a parallel degree of six, which forms nine groups. Considering practical task allocation and execution occupancy, not all groups can map to a contiguous physical ring. As exemplified, up to six groups (marked in red) result in non-contiguous, tetris-like ring patterns that bottleneck communication and reduce compute utilization. This issue becomes more severe for larger models, such as Llama-2, on larger-scale WSCs, as shown in Fig.~\ref{fig:TATP motivation}(c), where topology mismatch can reduce compute utilization by over 30\%. Therefore, realizing efficient parallel execution on wafers necessitates tight synergy with the physical topology to minimize communication overheads.

\begin{algorithm}[t]
\caption{Bidirectional Tensor Stream Orchestration}
\label{alg:1D orchestration}
\begin{algorithmic}[1]
\Require Number of dies $N$, die ID $src$, time step $t$;
\State Initially, $subT[src]$ is located on $die_{src}$;
\State Execute computation phase:
    \Statex \hspace{0.1em} { $\triangleright$ When $src < N/2$}:
    \State \hspace{1.5em} \texttt{Compute} using $subT[(src+t) \bmod N]$;
    \Statex \hspace{0.1em} { $\triangleright$ When $src \ge N/2$}:
    \State \hspace{1.5em} \texttt{Compute} using $subT[(src-t+N) \bmod N]$;
\State Concurrent communication phase:
    \Statex \hspace{0.1em} {\small $\triangleright$ When $t \leq src < N-1$}:
        \State \hspace{1.5em}  \texttt{Send}$(src, src-1, subT[(src+t) \bmod N])$; 
    \Statex \hspace{0.1em} {\small $\triangleright$ When $0 < src \leq N-t-1$}:
        \State  \hspace{1.5em} \texttt{Send}$(src, src+1, subT[(src-t+N) \bmod N])$; 

    \Statex \hspace{0.1em} {\small $\triangleright$ When $t > \frac{N}{2} -1 $ and $t - \frac{N}{2} \leq src < t$}:
        \State \hspace{1.5em} \texttt{Send}$(src, src-1, subT[(src-N+t) \bmod N])$; 
    \Statex \hspace{0.1em} {\small $\triangleright$ When $t > \frac{N}{2} -1 $ and $N-t < src \leq \frac{3N}{2}-t$}:
        \State \hspace{1.5em} \texttt{Send}$(src, src+1, subT[(N+src-t) \bmod N])$; 
\Ensure Coordinated Comp and Comm across dies;
\end{algorithmic}
\end{algorithm}

Fig. \ref{fig:Tensor flow partition}(a) illustrates the co-design flow of TATP, which integrates parallel-level and logical-level designs. At the parallel level, tensors are partitioned into fine-grained sub-tensors and a logical ring communication pattern is defined. At the logical level, TATP optimizes the execution schedule to align the logical ring with the physical topology, and feeds back two alternative tensor-transfer options to the parallel level, thereby enhancing communication–computation overlap.
\begin{equation}
\begin{aligned}
& \textbf{Forward:} \quad~~ \mathbf{O}_{[B,\,M,\,K]} = \mathbf{I}_{[B,\, M,\,N]} \times \mathbf{W}_{[N,\,K]}, \\
&\textbf{Backward:} \quad  \mathbf{dI}_{[B,\,M,\,N]}  = \mathbf{dO}_{[B,\,M,\,K]} \times \mathbf{W}^T_{[K,\,N]}, \\
&\textbf{Gradient:} \quad ~~ \mathbf{dW}_{[N,\,K]} = \mathbf{I}^T_{[B,\, N,\,M]} \times \mathbf{dO}_{[B,\,M,\,K]}.
\end{aligned}
\label{eq:matmul}
\end{equation}

Specifically, at the parallel level, a training step for a linear operator consists of three stages: \textit{Forward}, \textit{Backward}, and \textit{Gradient-update}, which are condensed in Eq.~\eqref{eq:matmul}. Following the partitioning strategy in \cite{song2020accpar,song2019hypar}, we assume that $\mathbf{O}$ and $\mathbf{dI}$ share the same partition to minimize tensor resharding overhead. For clarity, we use the forward stage to illustrate the TSPP partitioning design. The backward and gradient-update stages follow the same principle.

Fig.~\ref{fig:Tensor flow partition}(a) illustrates the forward pass executed on four dies. The input tensor $I$ and weight tensor $W$ are each split into four sub-tensors, which are co-located as $(I_i,W_i)$ on Die~$i$. Four rounds execute over a logical ring. Specifically, in round $r$, Die~$i$ computes $O_{i,(i+r)\bmod 4}$, while the required sub-tensor is streamed in, fully overlapping communication with computation. After four rounds, each die holds a unique output slice. The backward and gradient-update stages reuse the same execution pattern, achieving identical overlap and efficiency.

\begin{figure}[t]
\centering
\includegraphics[width=\linewidth]{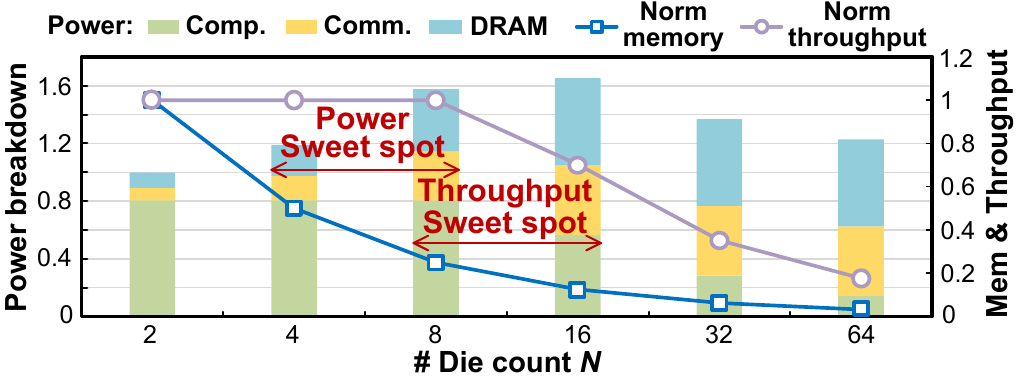}\vspace{-7mm}
\caption{Analysis of the sweet spot when implementing TATP on WSC.}
\label{fig:sweet spot of N}
\end{figure}

At the logical level, the ring communication requirement must be mapped onto the physical die array. As illustrated in Fig.~\ref{fig:Tensor flow partition}(a), in a 3$\times$4 die array with a parallel degree of four, not all three 4-die groups can be mapped to contiguous physical rings. For example, the group comprising Dies 0–3 forms a non-contiguous, tetris-like pattern rather than a physical ring. This mismatch between logical communication and physical topology forces multi-hop transfers, resulting in severe tail-latency. 


To resolve this topology-induced inefficiency, TATP introduces a logical‐level execution orchestration, as depicted in Fig. \ref{fig:Tensor flow partition}(b). A naive logical orchestration employs a direct ring communication pattern, which minimizes data transfer volume and is well suited for low-bandwidth links. However, such a design forces sub-tensor transfers to traverse $O$($N$) hops on WSCs, stalling computation and incurring severe tail latency. In contrast, TATP employs a \textbf{bidirectional redundant-transfer orchestration} that explicitly exploits the high-bandwidth D2D links of WSCs. Each sub-tensor is transmitted simultaneously in both directions while communication is carefully interleaved with computation. As a result, each die computes exactly one sub-output per round, and all data transfers traverse at most one physical hop, effectively eliminating long-tail latency.

For instance, in sub-weight streaming (Fig. \ref{fig:Tensor flow partition}(b)), the naive ring forces Die 3 to wait for a three-hop transfer of $W_0$ from Die 0 before it can compute $O_{30}$, introducing substantial tail latency that worsens as the die array scales. Instead, TATP’s relay-based orchestration delivers $W_2$, $W_1$ and $W_0$ in successive rounds via one-hop transfers from Die 2, enabling Die 3 to compute $O_{33}$, $O_{32}$, $O_{31}$, and $O_{30}$ without long-hop delays and to seamlessly overlap communication with computation.

Notably, TATP incorporates a selective transfer policy, under which the logical level determines whether sub-weights or sub-inputs are streamed during parallel execution. To minimize communication overhead, the policy consistently selects the smaller data type for transfer, which is particularly important for models with long sequences. For instance, in Llama2-7B with a sequence length over 14k, activations are approximately 3$\times$ larger than weight tensors. In this case, TATP prioritizes to transfer weights, alleviating communication burden. 

Taking sub-weight streaming as a exemple, TATP’s orchestration, illustrated in Fig. \ref{fig:Tensor flow partition}(c) and detailed in Alg.\ref{alg:1D orchestration}, consists of two tightly coordinated phases. During the computation phase (lines 2–4), each die processes exactly one sub-output per round, ensuring balanced workload distribution. For example, in Round 1, Dies 0–3 process $W_1$, $W_2$, $W_1$, and $W_2$, generating $O_{01}$, $O_{12}$, $O_{21}$, and $O_{32}$. After four rounds, the outputs match those of naive ring orchestration. The communication phase (lines 6–9) minimizes memory usage while ensuring each die receives the required sub-tensors on time. TATP uses a relay strategy to forward tensors for future use. For instance, in Round 0, Die 3 sends $W_3$ to Die 2 (line 6). While Die 2 uses $W_1$ in Round 1 to compute $O_{21}$ (line 4), $W_3$ is relayed through Die 2 to Die 1 (line 6), enabling Die 1 to compute $O_{13}$ in Round 2 (line 3).

\begin{figure}[t]
\centering
\includegraphics[width=\linewidth]{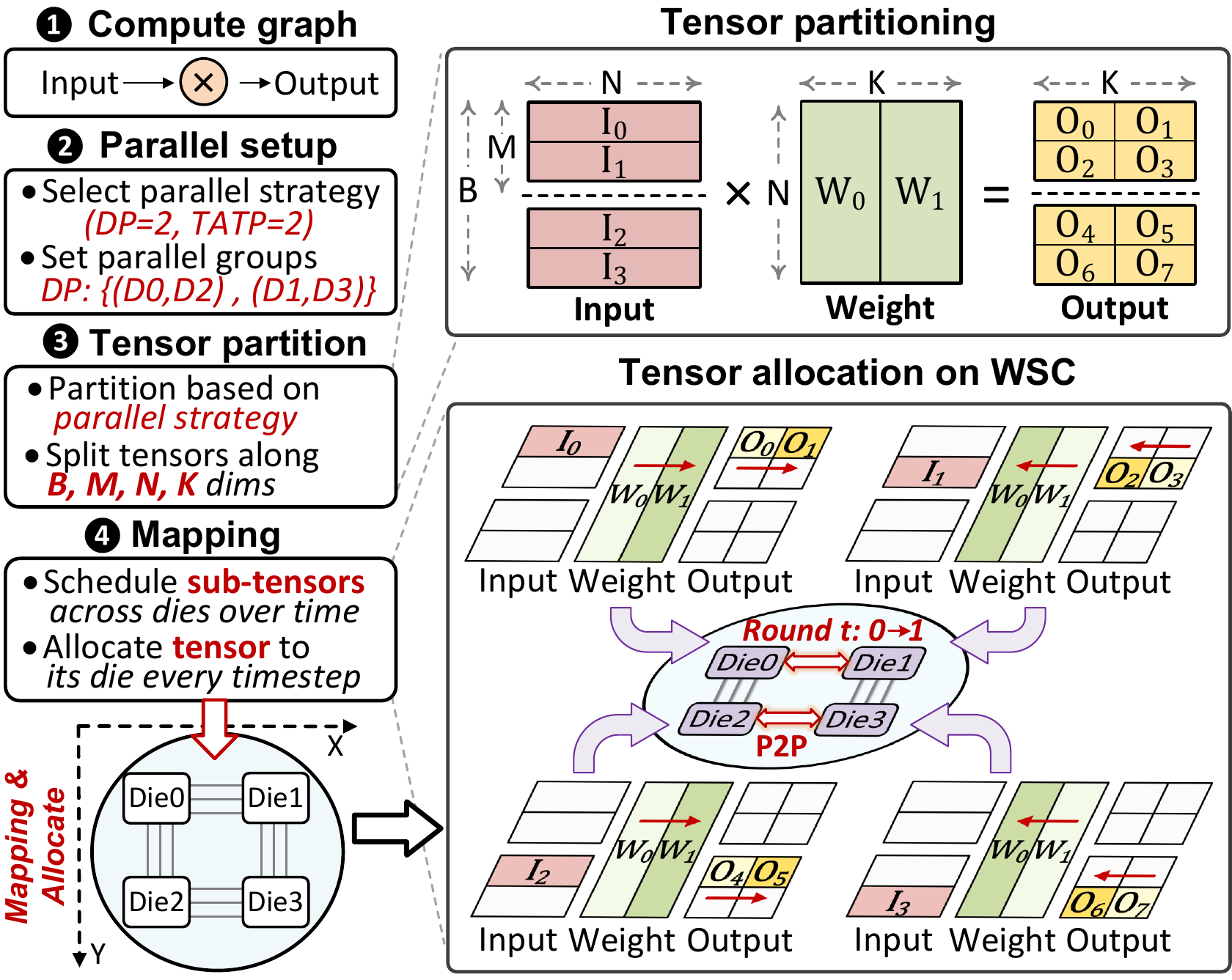}\vspace{-5mm}
\caption{Coordinate-based unified representation for hybrid parallelisms.}
\label{fig:mapping}
\end{figure}

\textbf{Insight: There is a key sweet spot of TATP parallel degree $N$ that achieves optimal throughput and power.} For a fixed workload (e.g., one GPT-3 175B linear layer) distributed across $N$ dies, per-die memory footprint and compute time scale as $O(1/N)$, whereas communication time remains constant (O(1)). As $N$ grows, compute tasks become finer-grained, causing communication time to exceed compute time, which stalls computation and becomes the bottleneck. As shown in Fig.~\ref{fig:sweet spot of N}, throughput and memory efficiency peak at $N\!\approx\!8$--$16$, then decline as communication overhead dominates. Interestingly, power consumption follows a different trend. The proportion of compute power stays flat until $N>8$, then drops; communication and DRAM power shares rise and then level off, producing a total power curve that first increases and then falls. Consequently, the optimal power-efficiency sweet spot sits at $N\!\approx$\,4-8 dies. In summary, choosing N in the range of roughly 4-16 dies delivers the best trade-off between throughput and power efficiency.

\section{TCME: Contention Ochestration}\label{sec:mapping engine}
In this section, building on TATP, we develop a Traffic-conscious Unified Mapping Engine (TCME) for WSCs to address the traffic contention arising from integrating TATP with existing parallel strategies and mapping them onto WSCs. TCME comprises of two key components: A \textit{unified parallelism representation} that enables precise identification of traffic contention, and a \textit{traffic-conscious communication optimizer} that systematically mitigates such contention.

\subsection{Unified Parallelism Representation} \label{sec:mapping}
The unified parallelism representation strategy takes a compute graph as input, applies hybrid parallelism for concurrent execution, and spatio-temporally maps the resulting parallel execution onto the WSCs. As illustrated in Fig. \ref{fig:mapping} \ding{182}, the example compute graph consists of a single linear operator. Taking a 2$\times$2 die array as an example, we configure hybrid parallelism strategies for the operator and map them onto the WSC. Any parallel strategy can be selected, including DP \cite{li2020pytorch}, FSDP \cite{zhao2023pytorch}, SP \cite{li2021sequence}, TP \cite{shoeybi2019megatron}, and TATP. Notably, the parallel degree of each strategy is set to match the die count, thereby forming the corresponding parallel groups on the WSC. For example, in Fig. \ref{fig:mapping} \ding{183}, both DP and TATP are configured with a degree of two on the four-die array. The DP groups are \{Die\,0, Die\,2\} and \{Die\,1, Die\,3\}, while the TATP groups are \{Die\,0, Die\,1\} and \{Die\,2, Die\,3\}.

Based on the selected parallel strategies, the unified parallelism representation 
strategy splits the input, weight and output tensors along the B (batch), M (sequence), N (hidden) and K (intermediate) dimensions. As examplified in Fig. \ref{fig:mapping} \ding{184}, DP splits the batch dimension B into two slices, while TATP splits the sequence and hidden dimensions M and K into two slices each, yielding four sub-inputs ($I_0$-$I_3$), two sub-weights ($W_0$, $W_1$) and eight sub-outputs ($O_0$-$O_7$). Here $I_1$ denotes the first batch slice and second sequence slice. We then define a spatio-temporal mapping that streams sub-tensors across dies over time and assigns each computation to a specific die in each round. As depicted in Fig. \ref{fig:mapping} \ding{185}, point-to-point (P2P) communication is temporally scheduled to transfer sub-weights between Die 0 and Die 1, as well as between Die 2 and Die 3. We spatially allocate sub-inputs $I_0$-$I_3$ and sub-weights in the pattern $W_0$, $W_1$, $W_0$, and $W_1$ to Die 0-3. In Round 0, Dies 0 and 1 compute $O_0$ and $O_3$ using $W_0$ and $W_1$, respectively, while simultaneously exchanging sub-weights. In Round 1, Die 0 holds $W_1$ and computes $O_1$, while Die 1 holds $W_0$ and computes $O_2$.

In summary, this unified parallelism representation strategy enables TCME to project the compute graph onto the WSC using hybrid parallel strategies, allowing precise identification of communication contention both across parallel strategies and among parallel groups.


\subsection{Traffic-conscious Communication Optimizer}\label{sec:comm optimizer}
\begin{figure}[t]
\centering
\includegraphics[width=\linewidth]{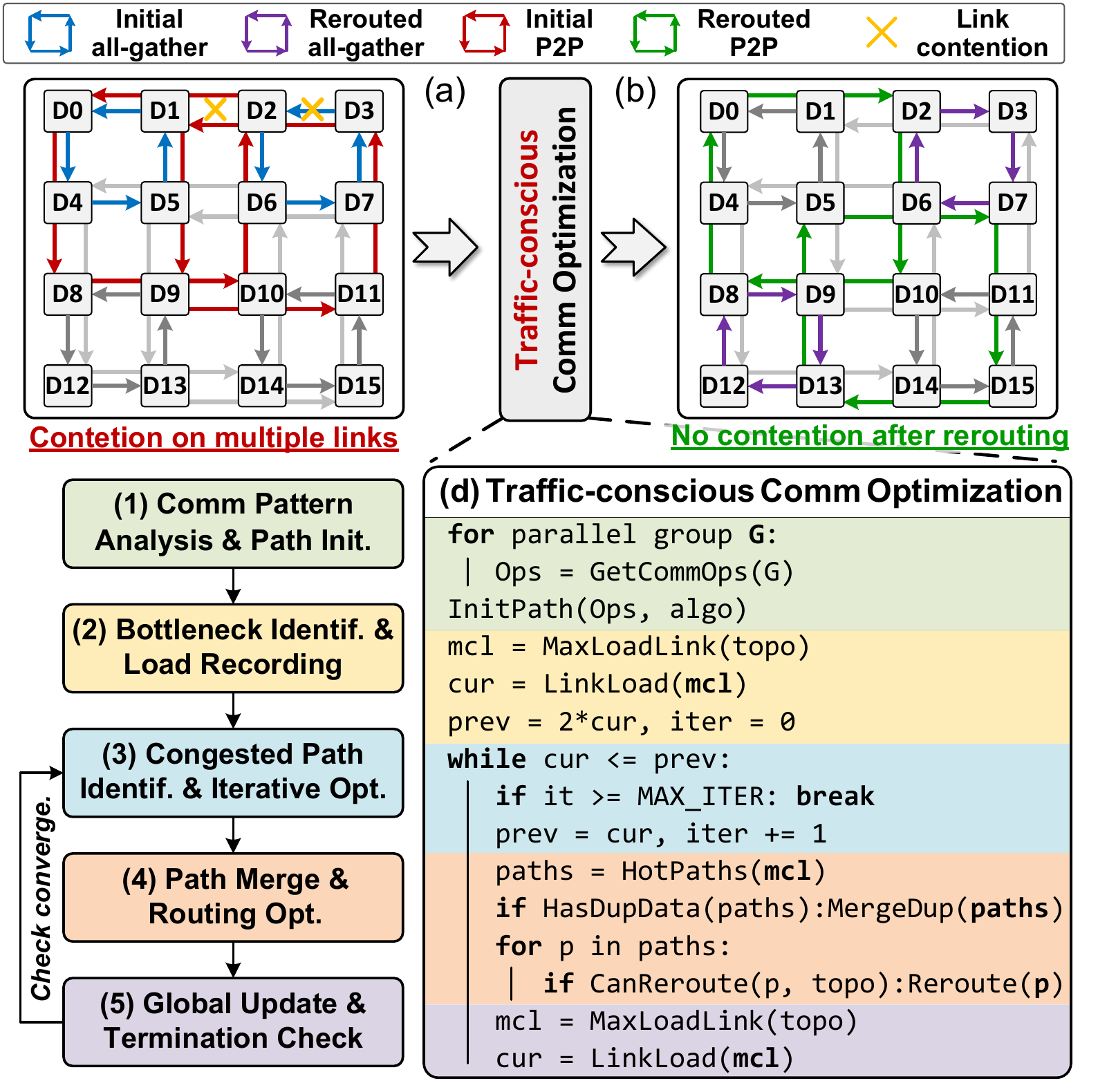}\vspace{-5mm}
\caption{Traffic-conscious communication optimizer in the mapping engine. (a) Example baseline communication pattern. (b) Example optimized communication pattern. (c) Flowchart overview. (d) Algorithm for the optimizer.}
\label{fig:comm optimizer}
\end{figure}

Building on the unified parallelism representation, TCME further addresses the severe traffic contention introduced by hybrid parallelism on WSCs, where multiple parallel groups contend for limited routing resources. Such contention is largely overlooked by existing frameworks, including Megatron \cite{narayanan2021efficient} and Gemini \cite{cai2024gemini}. As shown in Fig. \ref{fig:comm optimizer}(c) and (d), TCME incorporates a dedicated traffic-conscious communication optimizer that applies a five-phase workflow to coordinate dataflows and mitigate contention.


\textbf{(1)} \textit{Communication Pattern Analysis \& Path Initialization:} The optimizer first decomposes the hybrid parallel strategies to identify the set of parallel groups ($G$) and their associated communication operations ($Ops$). Based on these communication patterns, it initializes routing paths using standard communication algorithms \cite{awan2016efficient,awan2018optimized,hu2025demystifying}. However, as these algorithms are contention-agnostic, the resulting paths may lead to severe traffic contention, thereby motivating subsequent iterative refinement. 

\textbf{(2)} \textit{Bottleneck Identification \& Load Recording:} The optimizer performs a global analysis of execution traffic to identify the most congested link ($mcl$) and records its corresponding load ($cur$), which serves as the basis for subsequent contention mitigation. 

\textbf{(3)} \textit{Congested Path Identification \& Iterative Optimization:} The optimizer enters an iterative loop, identifying the set of congested paths ($paths$) that traverse the identified bottleneck link.

\textbf{(4)} \textit{Path Merging \& Routing Optimization:} Two complementary strategies are applied to optimize the congested paths ($paths$). Redundant path merging consolidates overlapping data flows into efficient multicast trees, while congestion-aware routing explores alternative bypass routes for the remaining paths.


\textbf{(5)} \textit{Global Update \& Termination Check:} The optimizer re-evaluates the global traffic state to identify the new most congested link $mcl$ and its corresponding load $cur$. The optimization loop terminates when load improvement stagnates or when the maximum iteration limit ($MAX\_ITER$) is reached, producing the final optimized communication patterns.

Fig.~\ref{fig:comm optimizer}(a) exemplifies a $4\times4$ die array, where both FSDP and TATP are configured with a parallel degree of 4. The Dies are labeled $D_0$–$D_{15}$. FSDP groups consist of four adjacent dies (e.g.\ $\{D_0,D_1,D_4,D_5\}$, $\{D_2,D_3,D_6,D_7\}$) and execute an all-gather of sub-weights in four rounds. This process, while contention-free within each group, continuously occupies links such as $Link_{2\rightarrow0}$. TATP groups span non-contiguous dies (e.g.\ $\{D_0,D_2,D_8,D_{10} \}$, $\{D_1,D_3,D_9,D_{11}\}$) and require a chain of P2P transfers (e.g., $D_2 \! \rightarrow \! D_0 \! \rightarrow \! D_8 \! \rightarrow \! D_{10}$). Since $Link_{2\rightarrow0}$ is held by the FSDP all-gather for all four rounds, any TATP transfer from $D_2$ to $D_1$ is inevitably blocked, creating severe contention on links like $Link_{2\rightarrow1}$ and $Link_{7\rightarrow3}$.

The optimizer identifies that contention arises from overlapping FSDP all‐gather and TATP P2P paths at both the intra-group and inter‐group levels. As exemplified in Fig.~\ref{fig:comm optimizer}(a), the all‐gather of FSDP requires traverse  $D_1$$\rightarrow$$D_0$$\rightarrow$$D_4$$\rightarrow$$D_5$, $D_3$$\rightarrow$$D_2$$\rightarrow$$D_6$$\rightarrow$$D_7$, $D_9$$\rightarrow$$D_8$$\rightarrow$$D_{12}$$\rightarrow$$D_{13}$ and $D_{11}$$\rightarrow$$D_{10}$$\rightarrow$$D_{14}$$\rightarrow$$D_{15}$. Meanwhile the P2P communication of TATP requires transfers of $D_2$$\rightarrow$$D_0$$\rightarrow$$D_8$$\rightarrow$$D_{10}$, $D_3$$\rightarrow$$D_1$$\rightarrow$$D_9$$\rightarrow$$D_{11}$, $D_6$$\rightarrow$$D_4$$\rightarrow$$D_{12}$$\rightarrow$$D_{14}$ and $D_7$$\rightarrow$$D_5$$\rightarrow$$D_{13}$$\rightarrow$$D_{15}$. These flows all contend on hot links such as $Link_{2\rightarrow0}$ and $Link_{3\rightarrow1}$, while links like $Link_{2\rightarrow3}$, $Link_{3\rightarrow7}$, $Link_{8\rightarrow0}$, and $Link_{0\rightarrow2}$ remain unused. By exploiting these idle links, the optimizer reroutes the all-gather paths from $D_3$$\rightarrow$$D_2$$\rightarrow$$D_6$$\rightarrow$$D_7$ and $D_9$$\rightarrow$$D_8$$\rightarrow$$D_{12}$$\rightarrow$$D_{13}$ to $D_2$$\rightarrow$$D_3$$\rightarrow$$D_7$$\rightarrow$$D_6$ and $D_8$$\rightarrow$$D_9$$\rightarrow$$D_{13}$$\rightarrow$$D_{12}$, keeping $D_1$$\rightarrow$$D_0$$\rightarrow$$D_4$$\rightarrow$$D_5$ and $D_{11}$$\rightarrow$$D_{10}$$\rightarrow$$D_{14}$$\rightarrow$$D_{15}$ unchanged. Similarly, it redirects the P2P paths from $D_2$$\rightarrow$$D_0$$\rightarrow$$D_8$$\rightarrow$$D_{10}$ and $D_7$$\rightarrow$$D_5$$\rightarrow$$D_{13}$$\rightarrow$$D_{15}$ to $D_0$$\rightarrow$$D_2$$\rightarrow$$D_{10}$$\rightarrow$$D_8$ and $D_5$$\rightarrow$$D_7$$\rightarrow$$D_{15}$$\rightarrow$$D_{13}$, leaving $D_3$$\rightarrow$$D_1$$\rightarrow$$D_9$$\rightarrow$$D_{11}$ and $D_6$$\rightarrow$$D_4$$\rightarrow$$D_{12}$$\rightarrow$$D_{14}$ unchanged. By eliminating both intra- and inter-group contention between FSDP and TATP, the optimizer reduces network congestion and yields the optimized communication pattern shown in Fig.~\ref{fig:comm optimizer}(b).

\begin{figure}[t]
\centering
\includegraphics[width=\linewidth]{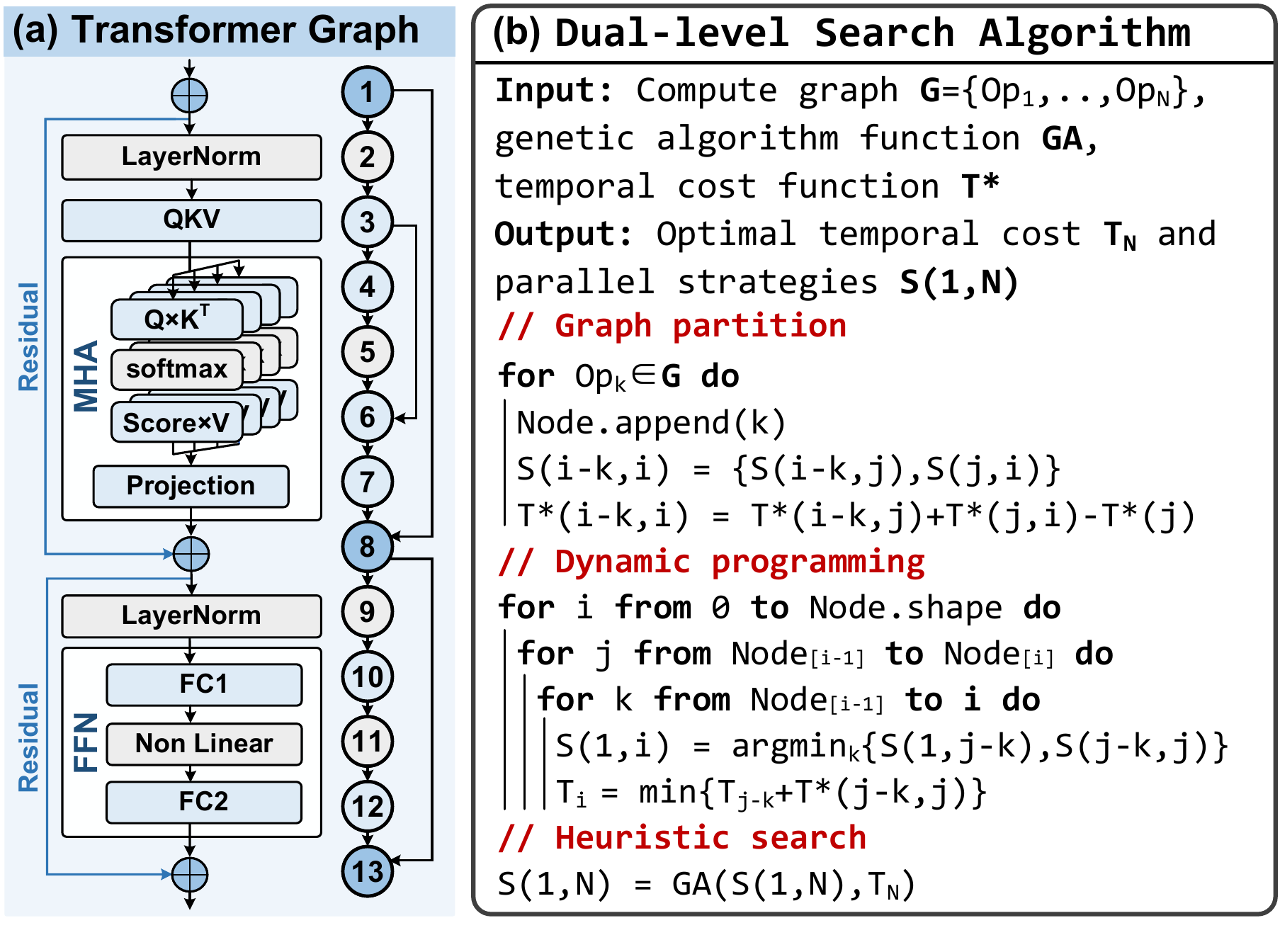}\vspace{-5mm}
\caption{(a) Transformer Architecture. (b) DLS algorithm.}
\label{fig:transformer block}
\end{figure}

\section{Dual-level Wafer Solver (DLWS)}\label{sec:wafer solver}
To accelerate the search time, we build a Dual-level Wafer Solver, which integrates a customized cost model with a dual-level search algorithm to efficiently generate rapid solutions.

\subsection{Customized Cost Model of WSCs}\label{sec:WSC cost model}
\noindent \textbf{{(1) Hardware level.}}\label{sec:hardware level}
We build upon ASTRA-Sim~\cite{won2023astra}, an open-source simulator validated against real hardware, to integrate our proposed TATP and TCME. Our extension models both computation and communication latencies, leveraging Ramulator~\cite{luo2023ramulator} to simulate memory occupancy. The simulation framework comprehensively covers essential computational operators such as GEMM, Softmax, GeLU, as well as inter-die communication primitives like P2P and collective algorithms~\cite{sanders2009two,tang2020communication,fang2024palm,awan2018optimized,hu2025demystifying}. For power consumption estimation, we calculate the total power as the sum of contributions from computing units, memory components, and communication interfaces.
The power consumption of each module is derived from its operational count and the corresponding energy cost per operation. For instance, the power consumed by the communication interface depends on both the data transmission volume and the link’s energy efficiency, with detailed parameters listed in Table~\ref{tab:wafer}.

To build a generalizable cost model, we first construct a comprehensive dataset using ASTRA-sim. This dataset spans a wide range of WSC configurations, including die array size, bandwidth, SRAM, network topologies (e.g., Mesh, Torus), and operator graphs. For each configuration, we profile key performance metrics, such as computation and communication latency, memory usage, and power consumption. This data is then used to train a DNN that learns the underlying relationships between system parameters and performance outcomes \cite{panner2024can, kim2024llmem, xu2025wsc}. The model generalizes to unseen workloads, such as novel Transformer variants, by parsing their unique operator dependencies and communication patterns from input features. Furthermore, its accuracy can be enhanced through fine-tuning on specific hardware configurations or workloads.

\textbf{(2) Compute graph level}. Fig. \ref{fig:transformer block} illustrates a typical Transformer block supported by TEMP. To optimize computational efficiency, TEMP integrates specialized operators, such as FlashAttention \cite{dao2022flashattention,wang2024sofa} and employs online softmax \cite{rabe2021self,lin2024infinite} to guarantee correctness, as shown in operators 4-7 in Fig. \ref{fig:transformer block}(a). The total cost of a compute graph, denoted as $T_{\rm total}(Graph)$, comprises both intra-operator and inter-operator costs, represented by $T_{\rm intra}(Op)$ and $T_{\rm inter}(Op_1,Op_2)$ respectively. The intra-operator cost includes computation latency, P2P communication, and collective communication overhead. The inter-operator cost, on the other hand, mainly involves P2P communication across dies. All these cost components can be expressed as functions of the operators. We can derive the expressions in Eqs. \eqref{eq:Intra_cost} and \eqref{eq:Inter_cost}:

\begin{table}[t]
\footnotesize
\renewcommand{\arraystretch}{1.05}
\caption{Waferscale Chip Configuration Parameters.}\vspace{-3mm}
\begin{center}
\begin{tabular}{l|cc}
\specialrule{0.12em}{0.5pt}{1pt}
\!\!\textbf{Modules} & \textbf{Parameters} & \!\!\textbf{Configurations}\!\! \\
\hline
\!\!\multirow{6}{*}{Logic Die}\!\! 
&\!\!Logic Die Area\!\! & 500mm$^2$\\
& \!\!SRAM Capacity \!\! &  80MB \\
& \!\!Die Router \!\! &  Input-queued Architecture \\
& \!\!Die-to-Die Interconnect \!\! & 4TB/s, 200ns, 5.0pJ/bit \\
& \!\!Operating Frequency \!\! & 2000MHz \\
& \!\!Computing Power \!\! &  \!\!1800TFLOPS, 2TFLOPS/Watt \!\! \\
\hline
\!\!\multirow{3}{*}{DRAM Die}\!\! 
&\!\! HBM Die Area \!\!& 210mm$^2$ \\
&\!\!HBM Capacity \!\! & 72GB \\
&\!\!Access Bandwidth\!\! & 1TB/s, 100ns, 6.0pJ/bit  \\
\specialrule{0.12em}{0.1pt}{0.1pt}
\end{tabular}
\end{center}
\label{tab:wafer}\vspace{-6mm}
\end{table}

\begin{equation}
\!\!T_{\text{intra}}(Op)\!=\!\text{Collective}(Op)+\max\{\text{Comp}(Op),\text{P2P}(Op)\}\!
\label{eq:Intra_cost}
\end{equation}

\begin{equation}
T_{\text{inter}}(Op_1,Op_2)=P2P(Op_1,Op_2).
\label{eq:Inter_cost}
\end{equation}

Accordingly, the total cost of the compute graph is derived as Eq. \eqref{eq:total cost}:

\begin{align}
T_{\text{total}}(Graph) = \!\!\!\!\!\!\!\sum_{Op_i \in Graph} \!\!\!\!\!\!\!\!T_{\text{intra}}(Op_i)  +\!\!\!\!\!\!\!\!\!\!\!\!\!\!\! \sum_{(Op_i, Op_j) \in Graph} \!\!\!\!\!\!\!\!\!\!\!\!\!\!T_{\text{inter}}(Op_i, Op_j).
\label{eq:total cost}
\end{align}

\subsection{Dual-level Search (DLS) Algorithm}\label{sec:dynamic algorithm}

Beyond the enormous search space illustrated in Fig. \ref{fig:motivation fig2}(c), the residual connections in Fig. \ref{fig:transformer block}(a) introduce additional optimization challenges. To address this, we propose a Dynamic Programming approach enhanced with Genetic Algorithm \cite{lambora2019genetic,peng2019generic} for rapid solution, as detailed in the dual-level search (DLS) Algorithm shown in Fig. \ref{fig:transformer block}(b).

For a compute graph with multiple operators, the DLS algorithm employs a dual-level search strategy, combining divide-and-conquer, recursive dynamic programming, and a genetic algorithm, to find the optimal strategy ${S_i}$ for each operator $Op_i$. First, it partitions the initial graph into $k$ sub-graphs with no residual connections, shrinking the search space from $O(N^2)$ to $O(N^2/k)$. At the first level, a recursive dynamic-programming routine iteratively optimizes one operator at a time, localizing decisions and drastically reducing complexity. At the second level, a genetic search refines those solutions by encoding the mapping engine’s parallel-setup parameters and spatio-temporal mappings as genes, then applying crossover, mutation, and elitist selection to evolve superior parallel strategies and tensor mappings. Because graph partitioning and dynamic programming have already pared down the search space, the genetic stage converges quickly, ensuring the DLS algorithm discovers high-quality solutions in minimal time.

\begin{table}[b]
\footnotesize
\renewcommand{\arraystretch}{1.1}
\caption{The Parameter Configurations of LLM Models.}\vspace{-5mm}
\begin{center}
\begin{tabular}{l|c|c|c|c|c}
\specialrule{0.12em}{0.5pt}{0.5pt}
\textbf{Model} & \textbf{Heads}
& \textbf{Batch} & \textbf{Hidden size} & \textbf{Layers}
& \textbf{Seq}\\
\specialrule{0.12em}{0.5pt}{0.5pt}
GPT-3 6.7B \cite{brown2020language} & 32 & 128 & 4096 & 32 & 2048 \\
\hline
Llama2 7B \cite{touvron2023llama} & 32 & 128 & 4096 & 32 & 4096 \\
\hline
Llama3 70B \cite{grattafiori2024llama} & 64 & 128 & 8192 & 80 & 4096 \\
\hline
GPT-3 76B \cite{narayanan2021efficient} & 80 & 128 & 10240 & 60 & 2048 \\
\hline
GPT-3 175B \cite{achiam2023gpt} & 96 & 128 & 12288 & 96 & 2048 \\
\hline
OPT 175B \cite{zhang2022opt} & 96 & 128 & 12288 & 96 & 4096 \\
\specialrule{0.12em}{0.5pt}{0.5pt}
\end{tabular}
\end{center}
\label{tab:llm-models}  
\end{table}

\begin{figure}[t]
\centering
\includegraphics[width=\linewidth]{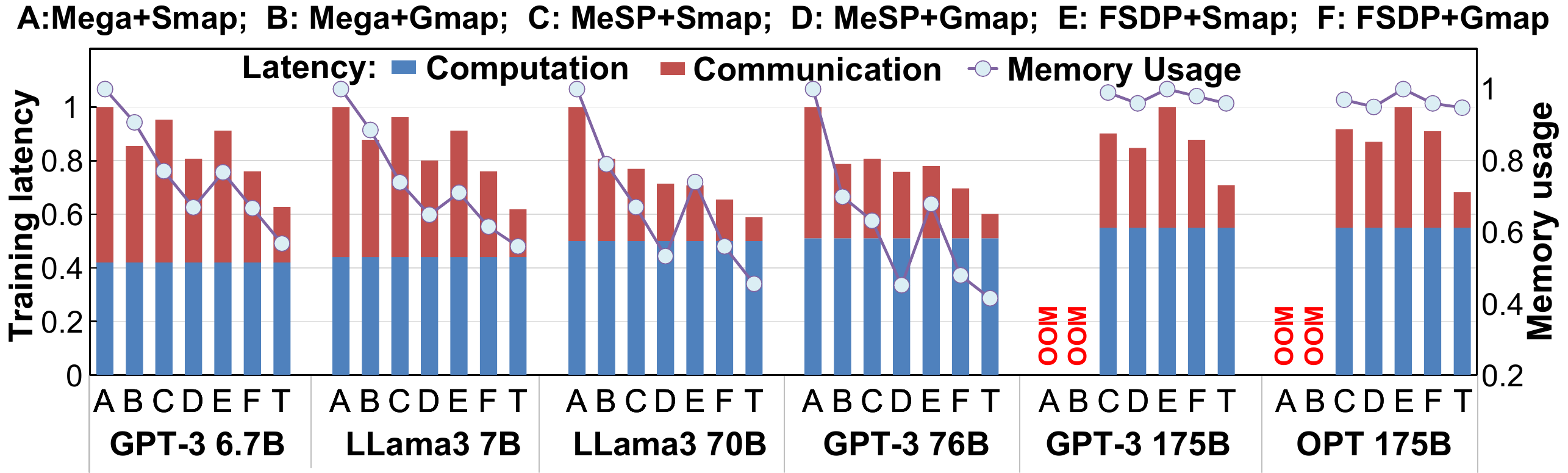}\vspace{-7mm}
\caption{Training performance comparisons for Mega+SMap, Mega+GMap, MeSP+SMap, MeSP+GMap, FSDP+SMap, FSDP+GMap, and TEMP.}
\label{fig:end to end throughput}
\end{figure}

\section{Evaluation}\label{sec:evaluation}
\subsection{Experimental Setup}\label{sec:experiment setup}

\noindent \textbf{WSC Configurations.} 
We select a WSC configuration with $4 \times 8$ dies operating at 2GHz. Each compute die, as shown in the configuration in Fig. \ref{fig:Hardware configurations of WSCs}, measures 33.25mm $\times$ 24.99mm and contains an $8 \times 8$ core array. The dies integrate logic components built on TSMC's 7nm process and DRAM on a 16nm process, with inter-die communication facilitated by a D2D interface offering 4 TB/s of bandwidth. A comprehensive list of configuration parameters is provided in Table \ref{tab:wafer}.

 \noindent \textbf{Baselines and Workloads.} 
We comprehensively evaluate our framework against six baselines, which are systematically constructed by combining three SOTA partitioning schemes with two distinct mapping engines. The partitioning schemes are based on leading parallel training methods: (1) \textbf{Megatron-1}\mbox{\cite{shoeybi2019megatron,narayanan2021efficient}}, a hierarchical mix of parallelisms that combines data (DP), tensor (TP), and pipeline (PP) parallelism; (2) \textbf{Megatron-3}\mbox{\cite{korthikanti2023reducing}} augmented with newer context (CP) and sequence (SP) parallelism; and (3) \textbf{FSDP}\mbox{\cite{li2020pytorch,zhao2023pytorch}}, a memory-efficient sharding strategy widely used in GPU systems. The mapping engines include: (1) \textbf{SMap}, a baseline sequential mapper with a fixed parallel strategy order; and (2) \textbf{GMap}, a WSC-adapted implementation of the Gemini mapper \cite{cai2024gemini}. While GMap can allocate DNN layers to specific dies with variable ordering and degrees of parallelism, it has two critical shortcomings for WSCs: it fails to explore the vast mapping space and lacks contention-aware optimization. This $3\times2$ construction yields our six baselines: \texttt{Mega+SMap}, \texttt{Mega+GMap}, \texttt{MeSP+SMap}, \texttt{MeSP+GMap}, \texttt{FSDP+SMap}, and \texttt{FSDP+GMap}. To evaluate WSC's performance, we benchmark six representative and widely used LLMs, including Llama2 7B \cite{touvron2023llama}, Llama3 70B \cite{grattafiori2024llama}, GPT-3 6.7B \cite{brown2020language}, GPT-3 76B \cite{narayanan2021efficient}, GPT-3 175B \cite{achiam2023gpt}, and OPT 175B \cite{zhang2022opt}. Their parameter configurations are detailed in Table~\ref{tab:llm-models}. We adopt mixed-precision training, utilizing FP16 for weights and activations, and FP32 for Adam optimizers.

\begin{figure}[t]
\centering
\includegraphics[width=\linewidth]{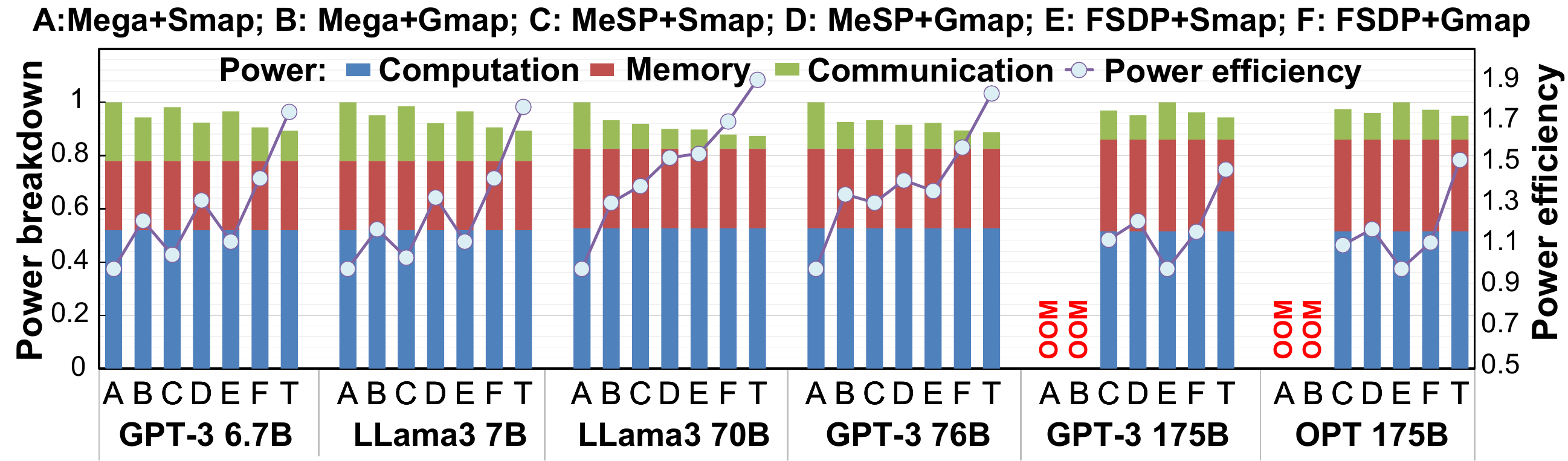}\vspace{-7mm}
\caption{Comparison of power efficiency among Mega+SMap, Mega+GMap, MeSP+SMap, MeSP+GMap, FSDP+SMap, FSDP+GMap, and TEMP.}
\label{fig:end to end power}
\end{figure}

\noindent \textbf{Simulator and Implementation.}
We build our simulation framework upon ASTRA-sim~\cite{won2023astra}, a precise open-source simulator that supports the modeling of wafer-scale computation and Network-on-Wafer (NoW) data routing. We further extend ASTRA-sim to implement the tensor-stream partitioning and communication optimizer developed in TEMP. For detailed memory simulation, we integrate the Ramulator~\cite{luo2023ramulator}, which provides fast and scalable DRAM modeling.

\noindent \textbf{Metrics.} 
We evaluate TEMP framework via four key metrics: \emph{training throughput} (computation efficiency and scalability), \emph{memory occupancy} (peak usage), \emph{bandwidth utilization} (communication efficiency), \emph{power efficiency} (throughput per Watt).

\subsection{Overall Performance Comparisons}\label{sec:overall throughput}

Fig.~\ref{fig:end to end throughput} compares the training performance of TEMP against various baseline strategies across multiple models. As shown, TEMP consistently outperforms all baselines, achieving end-to-end speedups of $1.69\times$ over \texttt{Mega+SMap}, $1.35\times$ over \texttt{Mega+GMap}, $1.38\times$ over \texttt{MeSP+SMap}, $1.24\times$ over \texttt{MeSP+GMap}, $1.39\times$ over \texttt{FSDP+SMap}, and $1.20\times$ over \texttt{FSDP+GMap}. The primary source of these performance gains is the substantial reduction in collective communication latency. Compared to the corresponding baselines, TEMP reduces collective communication latency by 38\%, 24\%, 27\%, 18\%, 25\%, and 14\%, respectively. 

These gains can be attributed to TEMP’s ability to address several fundamental limitations of existing approaches. \textit{FSDP}, despite its parameter sharding benefits, relies on coarse-grained all-gather communication patterns, which lead to severe network contention. \textit{Megatron’s hybrid parallelism}, while effective in reducing activation memory, still requires weight replication under data parallelism, resulting in unnecessary communication and memory overhead. In addition, the mapping strategies employed by existing systems exhibit inherent drawbacks. SMap enforces fixed priority rules, limiting its adaptability, whereas GMap lacks spatial awareness for WSC architecture, fails to coordinate mixed parallelism across layers, and does not optimize D2D communication. By jointly addressing these parallelization and mapping inefficiencies, TEMP achieves significantly lower communication overhead and superior end-to-end performance.  


\begin{figure}[t]
\centering
\includegraphics[width=\linewidth]{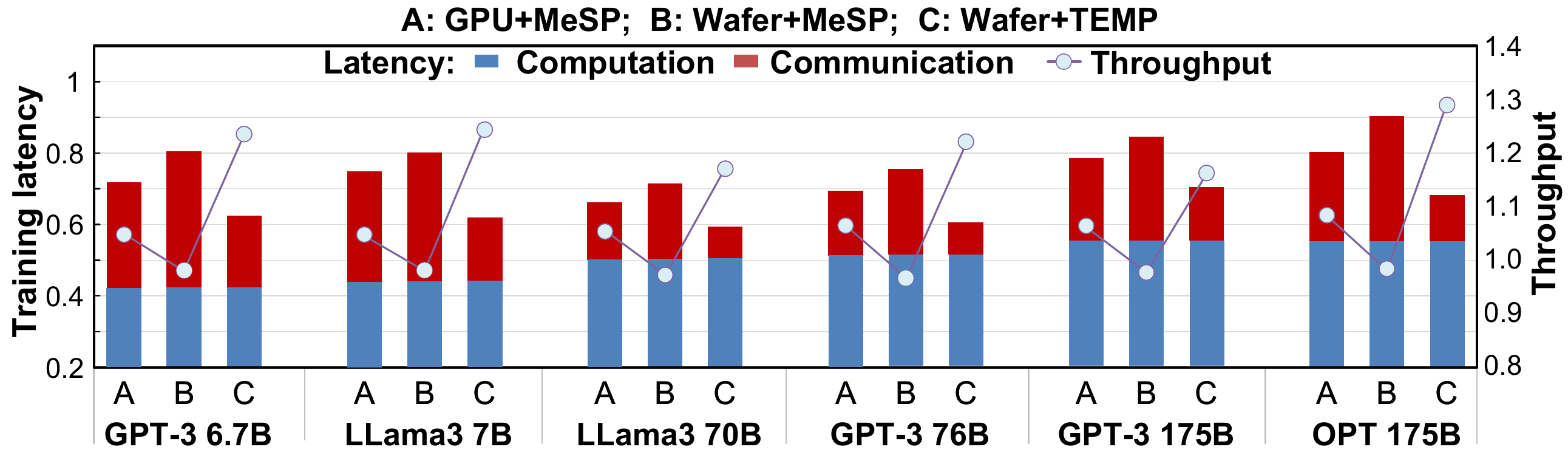}\vspace{-6mm}
\caption{Comparison of training performance among GPU cluster and WSC.}
\label{fig:gpu cluster}
\vspace{-2mm} 
\end{figure}

Regarding memory efficiency, we evaluate the peak occupancy on each method's best-performing configuration. As shown in Fig.~\ref{fig:end to end throughput}, TEMP consistently achieves the lowest peak memory, averaging just 49\% to 82\% of the usage of the baselines. Megatron-1 based approaches suffer from excessive tensor replication, leading to Out-of-Memory (OOM) errors on larger models. The varying memory performance can be explained by the underlying principles of different parallelism strategies. Both SP and TP operate on a similar idea: SP avoids replicating activations, while TP avoids replicating weights. For smaller models, where the memory footprints of activations and weights are modest, the memory-saving advantages of SP and TP are less pronounced. Instead, they introduce additional collective communication overhead. Consequently, the optimal strategy for \texttt{MeSP+SMap} and \texttt{MeSP+GMap} is to increase the degree of data parallelism (DP), which leads to higher memory consumption compared to TEMP. Conversely, for large models, relying heavily on DP can easily cause OOM errors. In this scenario, the memory-saving benefits of SP and TP become critical. This explains why, for the 175B model, the memory footprints of TEMP and the baseline methods are highly comparable, with differences falling within 10\%.

\textbf{Power Efficiency}. Fig.\mbox{~\ref{fig:end to end power}} further compares the power efficiency. As shown, TEMP's overall power consumption is 88.6\%, 94.5\%, 94.6\%, 98.6\%, 94.4\%, and 97.6\% of the respective baselines. This modest reduction is primarily attributed to a decrease in communication power, which TEMP lowers by 11\%, 5\%, 6\%, 3\%, 5\%, and 2\%, respectively. The fact that TEMP's overall power advantage is marginal is expected. As detailed in Table~\mbox{\ref{tab:wafer}}, computation is the dominant contributor to the system's total power consumption (over 50\%), given its high power draw compared to the more efficient D2D bandwidth power (5 pJ/bit). Since TEMP is not designed to optimize computation power, its impact on the total power consumption is naturally limited. Consequently, while total power savings are modest, power efficiency sees significant gains, mirroring the throughput improvements. TEMP achieves $1.85\times$, $1.45\times$, $1.47\times$, $1.23\times$, $1.48\times$, and $1.28\times$ higher power efficiency, respectively. This stems from TEMP's extensive search space exploration and communication refinement, which eliminates redundant data transfers to minimize interconnect energy consumption while maximizing throughput per Watt.

\begin{figure}[t]
\vspace{1.5mm}
\centering
\includegraphics[width=0.96\linewidth]{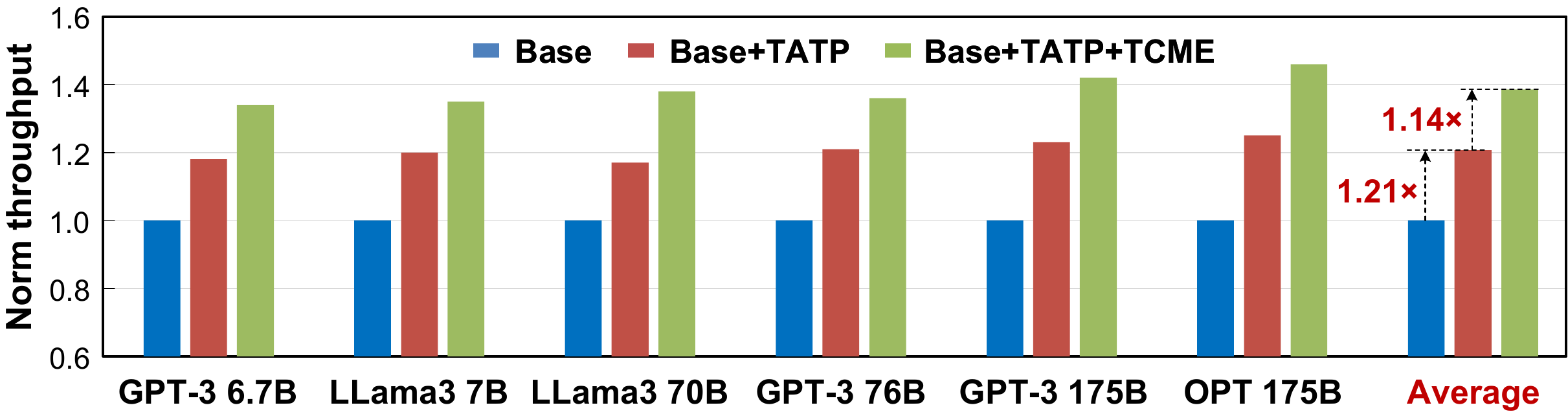}\vspace{-2mm}
\caption{Ablation performance of TEMP.}
\label{fig:ablation}
\vspace{-2mm} 
\end{figure}

\textbf{Comparison with GPU cluster}. To ensure a fair comparison of large model training performance between WSC and GPUs, we configure a 32-die WSC system to match the theoretical FP16 peak performance of a 4-node A100 GPU cluster (32 GPUs total, at 312 TFLOPS per GPU). We employ Megatron-3 for the GPU cluster, while applying MeSP+GMap and TEMP on the WSC system. The results, shown in Fig.~\ref{fig:gpu cluster}, indicate that \texttt{Wafer+TEMP} achieves the lowest training latency, delivering a speedup of $1.16\times$ over \texttt{GPU+MeSP} and $1.26\times$ over \texttt{Wafer+MeSP}, respectively. Notably, when both systems use a Megatron-3-like strategy, the GPU cluster exhibits better performance than the WSC system. This discrepancy stems from the inherent incompatibility of existing hybrid parallelism with the WSC architecture, leading to low utilization of its high-bandwidth interconnects. Conversely, by optimizing the WSC system with TEMP (which incorporates TATP and TCME), we can surpass the performance of the GPU cluster. This is because TEMP is specifically designed to fully leverage the unique high-bandwidth advantages of the wafer-scale fabric.

\subsection{Ablation Studies}

\begin{figure}[t]
\centering
\includegraphics[width=\linewidth]{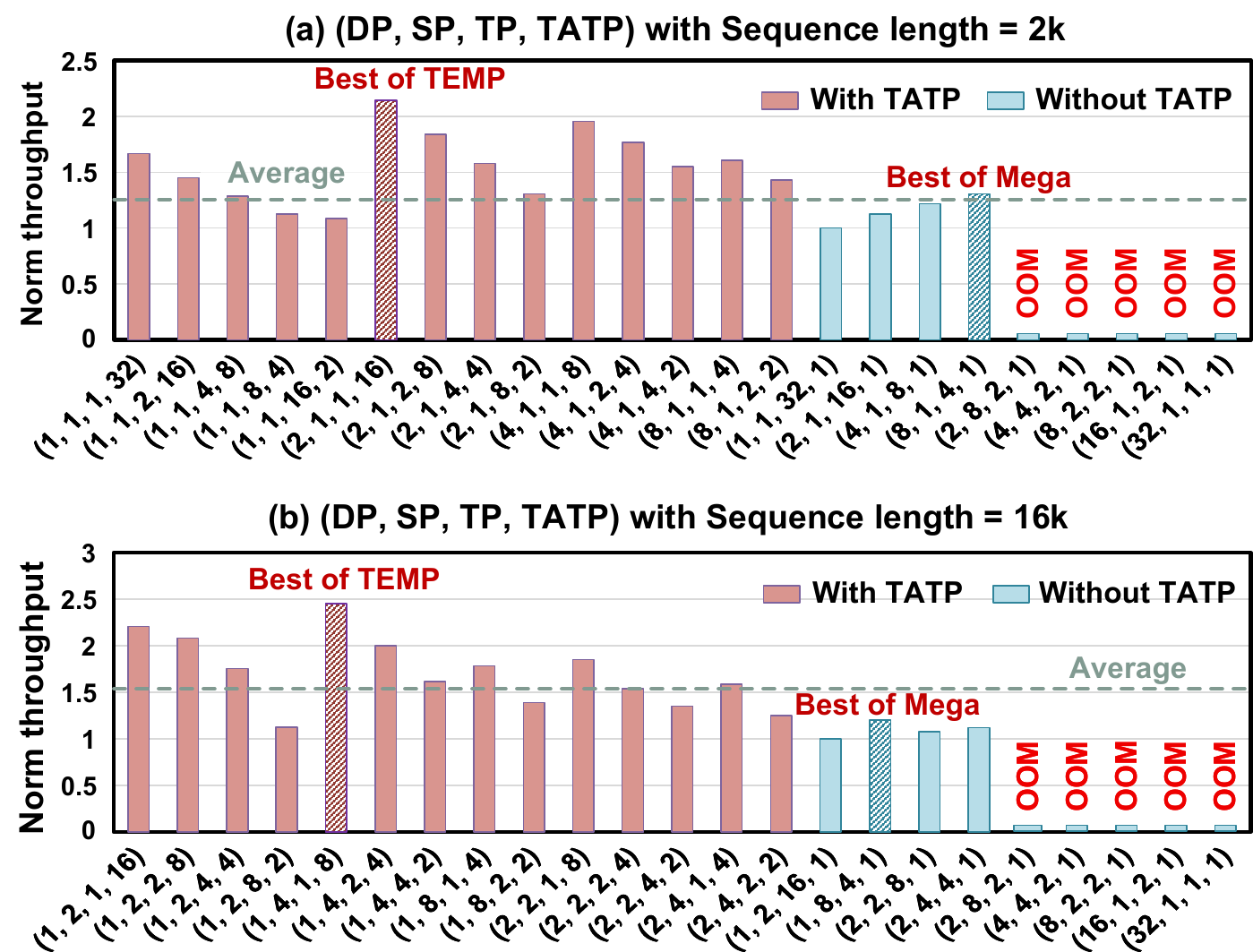}\vspace{-5mm}
\caption{Training performance of Llama2 7B with different parallel configurations (DP, TP, SP, TATP) using the same mapping engine TCME. (a) Batch size = 128, sequence length = 2k. (b) Batch size = 32, sequence length = 16k.}
\label{fig:partition granularity}
\vspace{-4mm}
\end{figure}

To evaluate the impact of individual optimization components in TEMP, we conduct an ablation study using training throughput as the primary metric. We select \texttt{FSDP+SMap} as the baseline configuration, as it is able to successfully train all benchmark models without encountering OOM errors.Starting from this baseline, we incrementally enable TEMP’s core optimizations: first enabling our Topology-aware Tensor-stream Partitioning (\textbf{+TATP}) and subsequently adding our Traffic-Conscious Mapping Engine (\textbf{+TCME}).

As shown in Figure~\ref{fig:ablation}, the results reveal two key trends. First, the performance gains from both TATP and TCME become more pronounced as the model size increases, highlighting TEMP's particular effectiveness for large-scale models. Second, TATP contributes a slightly greater performance benefit, improving throughput by an average of 1.21$\times$, compared to the 1.14$\times$ average from TCME. The primary reason for this is that as models grow, the memory and communication redundancy inherent in traditional parallelism becomes a more significant bottleneck. TATP's fine-grained partitioning directly mitigates this issue, and its communication overhead scales more gracefully. \textbf{These results indicate that by leveraging both TATP and TCME, TEMP supports high-performance LLM training across various models.}

\subsection{Analysis of Mixed-Parallelism Strategies}\label{sec:granularity}

\begin{figure}[t]
\vspace{3.5mm}
\centering
\includegraphics[width=\linewidth]{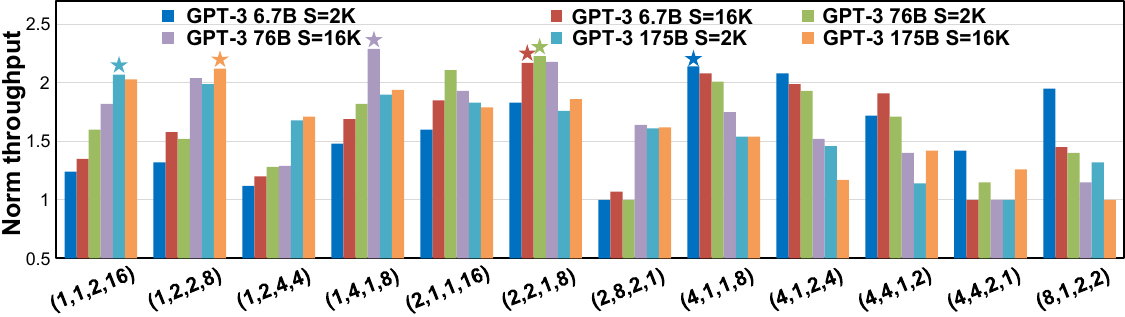}\vspace{-6mm}
\caption{Training throughput of GPT-3 models across various parallel configurations for short (S=2K) and long (S=16K) sequences. (1,1,2,16) denotes (DP=1, TP=1, SP=2, TATP=16).}
\label{fig:TATP converge}
\end{figure}

We evaluate partitioning granularity and mixed-parallelism strategies for LLM training using Llama2 7B on a 32-die wafer-scale chip. All configurations (DP, TP, SP/CP, and TATP) use our TEMP framework's TCME mapping engine for fair comparison. Performance with 2k and 16k sequence lengths is shown in Fig.~\ref{fig:partition granularity}(a) and (b). Two key observations can be derived from these results:

(1) Optimal performance requires a balanced TATP degree. Low degrees incur significant DP, TP, and CP overheads through tensor replication and collective communication, while high degrees create diminishing returns via fragmented workloads and resource underutilization. Fig.\ref{fig:partition granularity}(a) shows strategy (2,1,1,16) outperforming both (1,1,1,32) and (2,1,2,8), while Fig.\ref{fig:partition granularity}(b) demonstrates (1,4,1,8) surpassing alternatives. \textbf{These results highlight the essential synergy between TATP and existing parallelism strategies, proving hybrid approaches overcome limitations of individual methods.}

(2) TATP outperforms and synergizes with SP and CP, particularly for long sequences. TATP with degrees 8/16 consistently exceeds equivalent SP configurations by avoiding SP's high-overhead \texttt{All-Gather} operations and CP's weight replication requirements. Through fine-grained tensor streaming and topology-aware orchestration, TATP minimizes network latency while maximizing throughput. This creates a mutually beneficial relationship: SP helps TATP maintain optimal partitioning for computation-communication overlap, while TATP reduces SP's memory and communication costs. This explains why Fig.~\ref{fig:partition granularity}(b)'s top performer is the hybrid (1,4,1,8) configuration. \textbf{TATP in TEMP provides not merely memory and communication efficiency, but a topology-aware approach that complements DP, TP, SP, and CP to achieve superadditive performance.}

\textbf{Convergence of TATP:} To examine whether the optimal TATP dimension converges across training scenarios, we test three models (GPT-3 6.7B, 76B, and 175B) with short (S=2K) and long (S=16K) sequence lengths. Our results confirm that the optimal TATP dimension consistently falls within 8–16, as it maximizes the overlap of computation and communication under fixed D2D bandwidth, as the insight in Fig.~\mbox{\ref{fig:sweet spot of N}}. Fig.~\mbox{\ref{fig:TATP converge}} further validates this: regardless of model size or sequence length, the highest-throughput strategies consistently select a TATP dimension of 8 or 16. However, the optimal mix of DP, SP, and TP varies by scenario: DP dominates for short sequences, while SP and TP hybrids are superior for long sequences under high memory pressure. Key results include: (1) GPT-3 6.7B: Optimal configurations are (4,1,1,8) for S=2K and (2,2,1,8) for S=16K, with throughput gains of $2.12\times$ and $2.17\times$. (2) GPT-3 76B: Best strategies are (2,2,1,8) for S=2K and (1,4,1,8) for S=16K, achieving throughput improvements of $2.23\times$ and $2.29\times$. (3) GPT-3 175B: The best strategies are (1,1,2,16) for S=2K and (1,2,2,8) for S=16K, which result in throughput boosts of $2.06\times$ and $2.13\times$.

\subsection{Scalability of TEMP}
\begin{figure}[t]
\centering
\includegraphics[width=\linewidth]{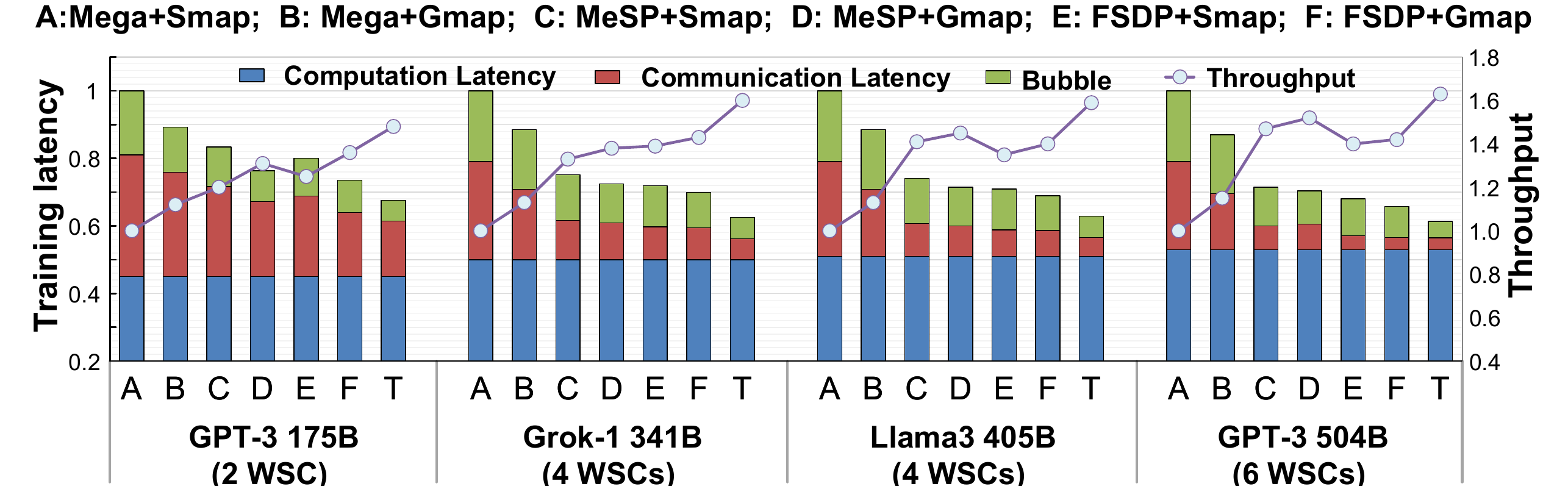}\vspace{-6mm}
\caption{Normalized training performance for large models on multi-WSCs.}
\label{fig:multiwafer}
\end{figure}

We further evaluate TEMP's scalability on a multi-wafer setup (\mbox{\S\ref{sec:experiment setup}}) using models like GPT-3 175B, Grok-1 341B, Llama3 405B, and a 504B GPT-3 variant. The number of wafers scales with model size: two for GPT-3 175B, four for Grok-1 and Llama3, and six for the 504B model, with pipeline parallelism (PP) used for inter-wafer model distribution. As shown in Fig.~\mbox{\ref{fig:multiwafer}}, TEMP achieves the highest throughput across all models, outperforming baselines by $1.2\times$ to $1.6\times$, thanks to its ability to reduce both intra-wafer communication time and pipeline bubbles. Baselines like \texttt{MeSP+GMap} and \texttt{FSDP+GMap} often resort to a high degree of pipeline parallelism, where the PP dimension is a multiple of the wafer count (e.g., PP=4 or 8 for the Grok-1 model on four wafers), which leads to large pipeline bubbles. This is a consequence of lacking parallelism strategies tailored for WSC, forcing them to choose from the standard DP, SP, and TP options. In contrast, TEMP introduces TATP, a novel parallelism method better suited for the long-distance communication patterns on a wafer. By enabling partial communication overlap, TATP allows for a lower PP dimension, which in turn reduces the pipeline bubbles. As depicted in Fig.~\mbox{\ref{fig:multiwafer}}, TEMP reduces the bubbles time by 14\%, 11\%, 6\%, 4\%, and 7\% compared to the respective baselines, while also cutting per-stage collective communication time by 22\%, 14\%, 7\%, 4\%, 5\%, and 3\%.

\subsection{Discussion on Fault Tolerance}
\begin{figure}[t]
\centering
\includegraphics[width=\linewidth]{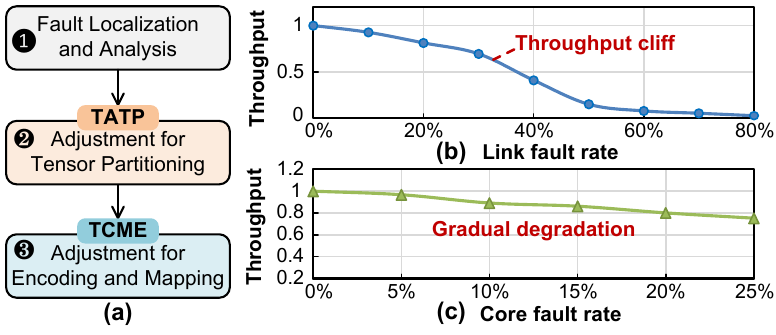}\vspace{-6mm}
\caption{Fault tolerance design and throughput exploration. (a) Fault tolerance mechanism integrated into TEMP. (b) Normalized throughput vs. link fault rate in WSC. (c) Normalized throughput vs. core fault rate in dies.}
\label{fig:fault tolerance} 
\end{figure}

We propose a systematic fault tolerance mechanism that surpasses hardware-centric methods. While systems like Cerebras \cite{Cerebras2022Cerebras} use redundant cores to maintain perfect mesh topology, they lack framework-level flexibility for imperfect meshes. Our approach adapts to hardware imperfections in large deployments through a three-step strategy (Fig.~\ref{fig:fault tolerance}(a)): (1) fault localization and classification, (2) adaptive tensor partitioning for computation re-balancing, and (3) communication re-routing around faulty hardware.

As shown in Fig.~\ref{fig:fault tolerance}(b) and (c), TEMP exhibits robust fault tolerance. While throughput is sensitive to link faults, hitting a performance cliff at a 35\% fault rate, the system is remarkably resilient to core faults. It sustains nearly 80\% of its peak throughput even at an extreme 25\% core fault rate. \textbf{These results underscore that TEMP's resilience is achieved through framework-level optimization, marking a fundamental departure from hardware-centric strategies that demand physical perfection.}

\subsection{Verification of DNN-based Cost Model}
To validate the accuracy of our DNN-based cost model, we adopt a multivariate regression model as the baseline and compare latency predictions obtained by different methods against simulation results across three categories: (1) Computation latency of single operator; (2) Communication latency for collective and point-to-point operations, and (3) Total latency considering computation-communication overlap. The computation operators include GEMM, GEMV, softmax, and SiLU, while the communication operators encompass All-Reduce, Reduce-Scatter, All-Gather, and P2P. For the overlap analysis, we focus on overlapping GEMM computation with the tensor communication mechanism employed in TATP.

\begin{figure}[t]
\centering
\includegraphics[width=\linewidth]{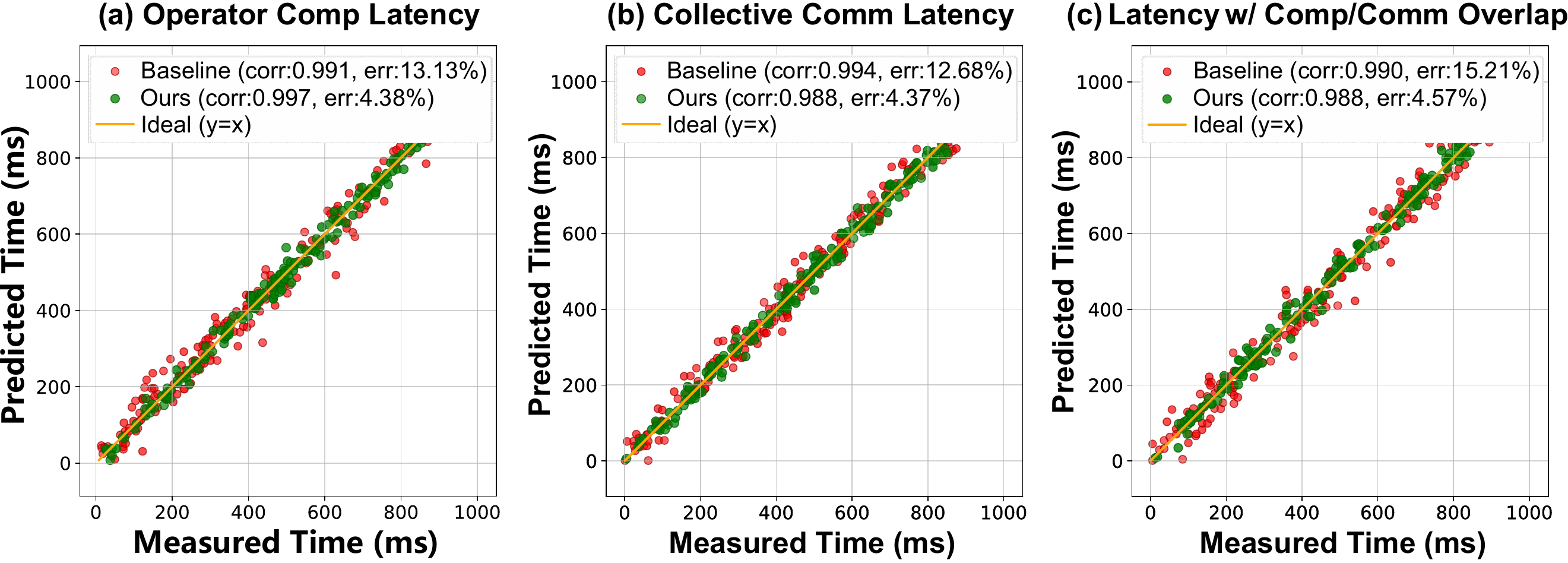}\vspace{-6mm}
\caption{The accuracy of DNN-based cost model for diverse metrics.}
\label{fig:DNN verification}
\end{figure}

By varying parameters such as batch size, sequence length, and hidden size, we generate 500 unique test cases. As shown in Fig. \ref{fig:DNN verification}, the baseline multivariate linear regression yields a correlation below 0.98 and an error rate reaching up to 10\%, reflecting limited predictive capability. In contrast, the DNN-based model demonstrates high fidelity, achieving correlations exceeding 0.99 and error rates of 4.38\%, 4.37\%, and 4.57\% for computation, communication, and overlap latency, respectively. These results confirm the model's accuracy. Furthermore, its lookup time is only a few hundred microseconds, significantly faster than simulation, which requires several minutes to over an hour, making the DNN-driven search process 100-1000$\times$ more efficient than simulation-based approaches.

\subsection{Search Time}\label{exp:search time}

We compare the execution time of our dynamic programming-based optimization algorithm with that of an integer linear programming (ILP) algorithm. All search time measurements are conducted on a single thread of an Intel Xeon Gold 5218 CPU (2.3GHz). 

For single-wafer configurations across the models evaluated in Section \ref{sec:experiment setup}, our optimization algorithm completes the search for the optimal configuration in approximately 3 minutes on average. In contrast, under the same model conditions, the ILP-based method \cite{zheng2022alpa} requires substantially longer execution time. Overall, our approach achieves a speedup of over 200× compared to the ILP baseline.


\subsection{Actionable Insights}

\textbf{Takeaway 1}: Across the trade-offs among throughput, memory efficiency, and power consumption, TATP achieves its best operating point at a parallelism degree of 8–16, indicating a robust and hardware-efficient design sweet spot.


\textbf{Takeaway 2}: The preferred parallel strategies vary based on model scale and sequence length. For smaller models (6B-70B), DP$+$TATP are favored for short sequences, while TATP becomes dominant for long sequences. For larger models (70B-200B), TATP$+$TP are prioritized for short sequences, while TATP$+$SP are the most effective for long sequences.

\textbf{Takeaway 3}: On multi-wafer systems comprising $N$ WSCs, introducing TATP fundamentally shifts the optimal parallelization strategy. Specifically, the optimal configuration transitions from high-degree pipeline parallelism (pp=$kN$) to a hybrid strategy with a reduced pipeline parallelism degree (pp=$N/k$). Across both intra-wafer and inter-wafer deployments, the optimal TATP parallel dimension consistently falls within 8 or 16. Notably, such cross-WSC deployment is practical due to the ample inter-wafer bandwidth (e.g., 9 TB/s \cite{talpes2022dojo}), which sufficiently supports the necessary communication and computation overlap.


\section{Related Work}\label{sec:related work}

\textbf{Wafer-scale Chips and Chiplets:} While recent research on WSCs has led to significant advancements in architecture \cite{feng2024switch,pal2021scale,chenwaferscale,rashidi2024fred,xu2025wsc,yang2025pd,pal2021designing,hu2024wafer,wang2024tmac,yu2025cramming}, floorplan design \cite{jiang2021cu,ozdemir2022kernel,luo2023ms,liu2022partition}, as well as the task scheduling and mapping \cite{wei2025spatial,li2024research,xu2025wsc}, the corresponding domain of WSC-contraint driven parallelization strategies remains relatively underexplored. Similarly, the literature on chiplets has primarily focused on design space exploration \cite{tan2024cocco, zhu2024theseus,cai2024gemini,orenes2024muchisim,iff2023rapidchiplet,huang2024hecaton,tan2021nn,hao2023monad} and dataflow optimization \cite{orenes2023dcra,wang2024task,graening2025chipletpart,li2025compass,cai2023inter}. However, these works often overlook the crucial role of interconnect topology, leaving a gap in methodologies for efficient, spatial-aware mapping and communication orchestration onto the physical hardware. TEMP pioneers the exploration of memory-efficient tensor parallelism under strict wafer-scale physical topology constraints.

\noindent \textbf{Tensor Partitioning and Distributed Training:} 
While significant progress has been made in tensor partitioning and distributed training for model parallelization \cite{jia2018exploring,jia2019beyond,knobe1990data,krizhevsky2014one,lepikhin2020gshard,song2023optimus,xu2021gspmd,zheng2022alpa,wang2024primepar,song2020accpar,song2019hypar,korthikanti2023reducing}, these works primarily target GPU systems with fully connected interconnection topologies, and thus fail to be effectively implemented on WSCs. For example, SP \cite{korthikanti2023reducing}, designed for long sequences, adheres to a tensor‑stationary paradigm that either leads to tensor replication or exposes communication to computation time.  While PrimePar \cite{wang2024primepar} realizes this challenge, and proposes spatial-temporal tensor partition to overlap communication and computation time, unfortunately, it requires diagonal D2D communication, which is more suited for GPU switch interconnects than for the 2D mesh interconnects used in WSCs. In contrast, TEMP accounts for the topology constraints of WSCs and utilizes a fine-grained tensor-stream dataflow along with holistic execution orchestration. This allows it to discover more efficient parallel strategies while avoiding interconnect congestion, resulting in improved memory and bandwidth efficiency.



\noindent \textbf{Mapping and Scheduling:} 
Numerous studies have focused on mapping and scheduling for tiled accelerators \cite{tan2024cocco,shao2019simba}, chiplets \cite{cai2024gemini,cai2023inter}, and GPUs \cite{zheng2020flextensor, sun2024adapipe}. However, these works largely overlook the physical arrangement of dies and the resulting communication patterns. This oversight limits their ability to develop truly topology‑aware parallelization strategies. Furthermore, some research \cite{tang2020communication,laskar2024enhancing,lim2025tidalmesh,laskar2025supermesh} has proposed specialized collective communication optimizations for 2D meshes, such as MultiTree \cite{tang2020communication}, TTO \cite{laskar2024enhancing}, and TidalMesh \cite{lim2025tidalmesh}. However, these methods are restricted to optimizing intra-group collectives i.e., communication within a single parallel group. In contrast, TEMP is explicitly designed to tackle the distinct challenge of inter‑group contention for physical links on WSCs, an issue that arises in hybrid parallelism. As such, TEMP complements these existing approaches by addressing a different aspect of the problem.

\section{Conclusion}\label{sec:conclusion}
We introduce TEMP, a memory-efficient and topology-aware framework designed for LLM training on WSCs. TEMP effectively addresses the challenges posed by topology constraints while optimizing both memory and bandwidth usage. It incorporates a unified parallelism representation and a dual-level search algorithm to identify optimal partitioning strategies.

\section*{Acknowledgment}
This work was supported in part by the National Science and Technology Major Project under Grant 2022ZD0115200; in part by the NSFC under Grant 62502255, Grant 62125403, Grant U24A20234, Grant 92464302 and Grant U24B20164; in part by the Beijing S\&T Project Z251100008425010; in part by Shanghai Municipal Science and Technology Major Project; the Natural Science Foundation of Jiangsu Province Basic Research Program under Grant BK20243042; in part by the Beijing National Research Center for Information Science and Technology; in part by the Northern IC Technology Innovation Center (Beijing) Co., Ltd under Grant QYJS20232801B; and in part by the Beijing Advanced Innovation Center for Integrated Circuits.

\bibliographystyle{IEEEtranS}
\bibliography{main}

\end{document}